\documentclass[twoside]{elsart}
\usepackage{fleqn}
\usepackage{graphicx}
\usepackage{amssymb}
\usepackage{amsmath}

\begin{document}

\begin{frontmatter}

\title{Direct Simulation of Low-Pressure Supersonic Gas Expansions and its 
Experimental Verification}
\author{Alexander Na{\ss} and Erhard Steffens}
\address{Physikalisches Institut, Universit\"at 
Erlangen-N\"urnberg, 91058 Erlangen, Germany}

\begin{abstract}
The use of gas expansions to generate atomic or molecular beams has become a 
standard technique in nuclear and hadron physics for the production of
polarized ion beams and gas targets. 
A direct simulation Monte Carlo method was used 
to understand the processes occurring in an expansion of highly dissociated 
hydrogen or deuterium gas at low densities. The results were verified in 
several measurements including time-of-flight and beam-profile determinations 
which showed that the supersonic gas expansions can properly be described by 
the Monte Carlo calculations. Additionally a new method of beam formation, the 
hollow carrier jet, was tested under the conditions of the atomic beam source 
operation.
\end{abstract}
\maketitle

\end{frontmatter}

\section{Introduction}
\label{intro}
Polarized atomic beam sources (ABS) as described, e.g., in 
Refs.~\cite{Haeberli_1967} and \cite{Iannotta_in_Scoles} are utilized to 
provide nuclear-polarized atomic hydrogen (H) and deuterium (D) beams. 
Molecular H$_2$ or D$_2$ gas is dissociated and the essentially atomic gas 
then expands through a cooled nozzle into the vacuum. A beam of high 
brightness is then formed by a skimmer and a collimator of  dimensions and 
positions to adapt the beam to the relatively small acceptance of the 
subsequent system of sextupole magnets. Based on the Stern-Gerlach principle, 
these magnets focus (defocus) atoms with electron-spin projection +1/2 (-1/2) 
along the magnetic field within the magnet bores. The electron-spin polarized 
beam then enters an rf transition unit, which allows to change the nuclear 
polarization by inducing transitions between the hyperfine states. 
Descriptions of the HERMES polarized ABS are given in 
Refs.~\cite{Stock_et_al_1994} and \cite{Nass_et_al_2003}. Details about its 
13.56\,MHz rf dissociator are found in Ref.~\cite{Stock_et_al_Koeln_1995}, 
whereas the 2.45\,GHz microwave dissociator and the atomic beam test stand are 
described in Ref.~\cite{Koch+Steffens_1999}.

To achieve a high output intensity, the atomic beam generated by the expansion 
has to fulfill several requirements like high flow rate, low transversal and 
longitudinal temperature, and a high degree of dissociation. The latter can 
only be achieved, if the recombination is low. This request sets a limit to 
the gas pressure in the dissociator volume of a few mbar. The pressure within
and at the exit of the nozzle corresponds to the transition region between 
laminar and molecular flow. There, the use of continuum-flow models (e.g., 
those based on the Navier-Stokes equations) is of restricted validity. Thus, a 
direct simulation Monte-Carlo (DSMC) method~\cite{Bird} was used to describe 
the processes during gas expansion. Time-of-flight (TOF) and, with the use 
of a novel type of monitor~\cite{Vassiliev_et_al_PST99}, beam-profile 
measurements were performed to check the validity of the results of the 
simulations. The investigation of the thermal properties of the gas in the 
nozzle region gave new insights. The achieved results could be used to improve 
the beam formation and to properly determine the beam parameters. The use of 
an over-expanded carrier jet, surrounding the inner atomic beam, had been 
proposed~\cite{Varentsov_et_al_Urbana_1997} to further increase the atomic 
beam intensity. This method of beam formation has been studied experimentally 
for the first time. The results could be interpreted by DSMC calculations. A 
detailed description of the performed work is found in 
Ref.~\cite{PhD_Nass_2002}.

\section{Supersonic Gas Expansion}
\begin{figure}[!b]
\centerline{\includegraphics[height=4cm,angle=0]{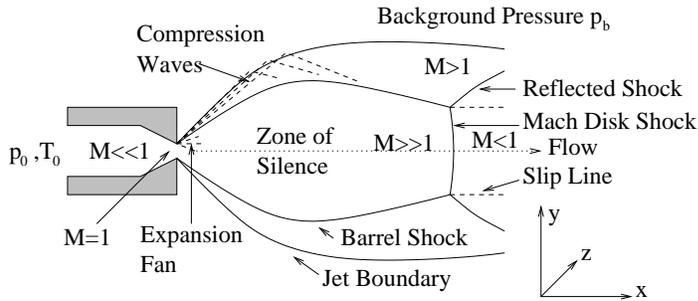}}
\caption{Continuum free-jet expansion of a gas into a region of background
pressure $p_{\rm b}$ starting from a negligibly small velocity 
at the stagnation state, described by $p_{\rm 0}$ and $T_{\rm 0}$ 
(figure taken from Ref.~\cite{Miller_in_Scoles}).}
\label{expan}  
\end{figure}

A free-jet atomic or molecular beam can be produced by a supersonic gas
expansion from a high-pressure gas source into a low pressure
background. Fig.~\ref{expan} shows the structure of a free expansion under
continuum (steady state) conditions.
The source is a short conical nozzle. As a result of the pressure difference 
$p_0-p_{\rm b}$, the gas is accelerated. At the nozzle exit the flow may 
reach sonic speed, if the ratio $p_0/p_{\rm b}$ exceeds the critical 
value~\cite{Miller_in_Scoles}
\begin{equation}
\label{eqg}
G=\left(\frac{\gamma+1}{2}\right)^{\gamma/(\gamma-1)}.
\end{equation}
From the possible values of $\gamma$, defined as 
$\gamma=c_{\rm p}/c_{\rm V}$ with the heat capacities $c_{\rm p}$ and 
$c_{\rm V}$, it follows that for all gases $G$ is lower than 2.05.
For $p_0/p_{\rm b}<G$ the flow exits subsonically. A supersonic beam has two 
important characteristic properties. First, the velocity of the beam $v$ 
increases during the expansion. Second, the beam parameters in the zone of 
silence (Fig.~\ref{expan})
are independent of boundary conditions (walls, $p_{\rm b}$), which is 
caused by the fact that information propagates at the speed of sound, whereas 
the gas moves faster. Therefore a skimmer is placed inside this zone to 
extract a supersonic beam. If the background pressure $p_{\rm b}$ is small 
enough, a smooth transition to molecular flow occurs and no shock structures 
emerge. The beam is only affected by residual gas scattering.

\subsection{Thermodynamic analysis}
Considering an ideal expanding gas without viscous and heat-conduction effects,
the energy equation~\cite{Miller_in_Scoles}
\begin{equation}
\label{eq1}
h+\frac{v^2}{2} = h_0
\end{equation}
holds, where $h_0$ is the total or stagnation enthalpy per unit mass and $v$
is the mean velocity in beam direction. For ideal gases 
(${\rm d}h = c_{\rm p} {\rm d}T$) and constant heat capacity 
$c_{\rm p} = (\gamma/(\gamma-1))(k_{\rm B}/m)$ one 
gets the maximum or terminal velocity (for $T \ll T_0$ after the expansion)
\begin{equation}
v_{\infty}=\sqrt{\frac{2k_{\rm B}}{m}\left(\frac{\gamma}{\gamma-1}\right)T_0} ,
\label{vinfty}
\end{equation}
with the particle mass $m$ and the Boltzmann constant $k_{\rm B}$. For H$_2$ 
($\gamma=5/3$) at $T_0=100$\,K, $v_{\infty}$ is 1436\,m/s. For mixtures of 
ideal gases, an average heat capacity 
\begin{equation}
\overline{c_{\rm p}}=\frac{\sum_i k_{\rm B} x_i \frac{\displaystyle \gamma_i}
               {\displaystyle \gamma_i-1}}{\sum_i x_i m_i}
\end{equation}
can be used, where $x_i$ is the fraction of the respective species. In the 
continuum limit the mean velocities of the species tend to be the 
same~\cite{Miller_in_Scoles}.

\begin{figure}
\centerline{\includegraphics[height=6cm,angle=0]{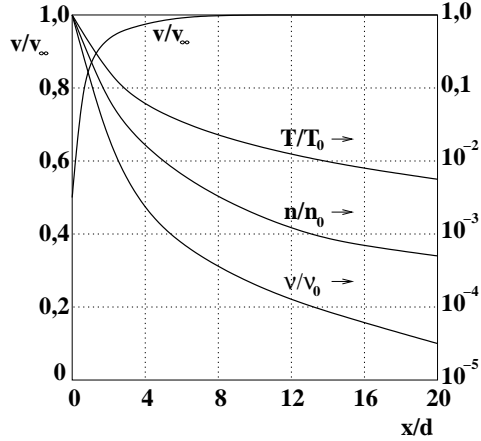}}
\caption{Free-jet on-axis properties versus distance from the source (given in 
source diameters) for a monoatomic gas ($\gamma$=5/3). The temperature $T$, 
the density $n$, and the binary hard-sphere collision frequency $\nu$ are
normalized by the source-stagnation values $T_0$, $n_0$, and $\nu_0$ (figure
taken from Ref.~\cite{Miller_in_Scoles}).}
\label{groessen}
\end{figure}

For isentropic expansion of an ideal gas, eq.~\ref{eq1} allows to deduce
\begin{equation}
\frac{T}{T_0}=\Big(1+\frac{\gamma -1}{2} M^2\Big)^{-1}
\end{equation}
with the assumption of constant $c_{\rm p}$. Here $M=v/c$ is the Mach number 
with 
the speed of sound $c=\sqrt{\gamma k_{\rm B} T/m}$. Furthermore, the mean beam 
velocity of the expanding gas as function of the Mach number is derived as 
\begin{equation}
v = M \sqrt{ \frac{\gamma k_{\rm B} T_0}{m}} \left(1+\frac{\gamma-1}{2}
  M^2\right)^{-1/2} .
\end{equation}
With the assumption $M=1$, which is expected around the nozzle exit, for H$_2$ 
$v$ is 718\,m/s. In general, with the use of $M$ as 
calculated~\cite{Miller_in_Scoles} for axisymmetric expansion, all 
thermodynamic variables for the free-jet expansion can be given as shown in 
Fig.~\ref{groessen} for their dependences along the centerline of the 
expansion.

Energy and enthalpy considerations~\cite{Haberland_et_al_1985}, based on 
Eqn.~\ref{eq1}, for an expanding beam of a monoatomic gas 
($c_{\rm p}=(5/2) (k_{\rm B}/m)$), lead to the balance equation
\begin{equation}
\frac{5}{2} k_{\rm B} T_0 =\frac{1}{2} m v_{\rm x}^2 + \frac{3}{2}k_{\rm B} 
T + k_{\rm B} T.
\label{E_definition}
\end{equation}
In the term on the left side $T_0$ is the stagnation temperature of the gas in 
the source, i.e., for thermal equilibrium the nozzle temperature 
$T_{\rm nozzle}$. The sum of the first and second term on the right side 
gives the total beam energy $E_{\rm beam}$ after transition to molecular
flow. The third term is the energy $k_{\rm B} T=p V$, stored in the gas and 
forcing it to expand. For $T_0=T_{\rm nozzle}$ one expects
\begin{equation}
\frac{3}{2} < \frac{E_{\rm beam}}{k_{\rm B} T_{\rm nozzle}} < \frac{5}{2}.
\label{E_boundaries}
\end{equation}

\subsection{Statistical definitions for the Monte Carlo method}
The Monte Carlo method is another way to obtain the parameters of the
atomic or molecular beam by simulating binary hard-sphere collisions between
the particles of the expanding gas. The thermal velocity ${\bf v'}$ of a 
particle of velocity ${\bf v_{\rm p}}$ in a beam of mean velocity ${\bf v}$ is 
defined by
\begin{equation}
\bf v_p' = v_p - v.
\end{equation}
The scalar pressure is defined as~\cite{Bird}  
\begin{equation}
p = \frac{1}{3} n m (\overline{v_{\rm p}'^2}),
\end{equation}
where $n$ is the particle density, $m$ the particle mass, and 
$v_{\rm p}'= \vert {\bf v_{\rm p}'}\vert$. With the ideal gas law 
$p=nk_{\rm B}T$, the average kinetic energy associated with the translational 
motion becomes
\begin{equation}
E_{\rm tr} = \frac{1}{2} m \overline{v_{\rm p}'^2} = 
\frac{3}{2} k_{\rm B} T_{\rm tr} .
\end{equation}
For every component $j$ this can be written as
\begin{equation}
m \overline{v_{{\rm p},j}'^2} = k_{\rm B} T_{{\rm tr},j} , \quad j = x, y, z
\label{eqtemp}
\end{equation}
For molecules the rotational and vibrational excitations have to be considered.
They can be ascribed to an internal energy
\begin{equation}
E_{\rm int} = \frac{1}{2} \zeta k_{\rm B} T_{\rm int} ,
\end{equation}
where $\zeta$ is the number of internal degrees of freedom.

For a gas in (local) thermal equilibrium, the fraction of the particles which 
are found within a velocity space element d${\bf v_p}$ is given by the thermal 
velocity distribution~\cite{Bird}
\begin{eqnarray}
\nonumber \frac{{\rm d}n}{n}&=&(\frac{\beta}{\sqrt{\pi}})^3
\exp(-\beta^2 v_{\rm p}'^2){\rm d {\bf v_p'}}\\
     &=&(\frac{\beta}{\sqrt{\pi}})^3\exp[-\beta^2(v_{\rm p,x}'^2
+v_{\rm p,y}'^2+v_{\rm p,z}'^2)]{\rm d}v_{\rm p,x}' 
{\rm d}v_{\rm p,y}' {\rm d}v_{\rm p,z}',
\end{eqnarray}
where $\beta = \sqrt{m/(2k_{\rm B}T_{\rm tr})}$. The fraction of particles 
with a velocity component in $j$ direction within the velocity range 
$v'_{{\rm p},j}$... $v'_{{\rm p},j}$+d$v'_{{\rm p},j}$, irrespectively of the 
magnitude 
of the other components, is obtained by integrating over the two other 
components. The distribution function for the thermal velocity component then 
is
\begin{equation}
\label{maxeq}
f(v_{{\rm p},j}') = \frac{\beta_j}{\sqrt{\pi}} \exp[-\beta_j^2 
v_{{\rm p},j}'^2]
=\frac{\beta_j}{\sqrt{\pi}} \exp[-\beta_j^2 
(v_{{\rm p},j}-v_{j})^2],
\end{equation}
where $\beta_j = \sqrt{m/(2k_{\rm B}T_{\rm tr,j})}$.
The most probable thermal velocity of each component is zero.

\section{Simulation Program and Experimental Setup}
\subsection{The direct simulation Monte Carlo program}
The direct simulation Monte Carlo method (DSMC)~\cite{Bird} is a technique for
the computer modeling of a real gas by some thousands or millions of
simulated particle trajectories. The velocity components and position 
coordinates of these
particles evolve in time as the particles are concurrently followed through 
representative collisions and boundary interactions in the physical 
space. The decoupling of the motion and collisions of the particles over small 
time steps and the division of the flow field into small cells are the key 
computational assumptions associated with the DSMC method. The time step 
should be much smaller than the mean collision time and a typical cell 
dimension should be much smaller than the local mean free path. The 
program~\cite{DSMC} has a flexible system for the specification of the flow 
geometry. For our purpose the geometry of the beam forming elements nozzle, 
skimmer, and collimator are implemented as boundary walls with temperature 
$\rm T$ for an axially symmetric flow. 
\begin{figure}[h]
\centerline{\includegraphics[height=5cm,angle=0]{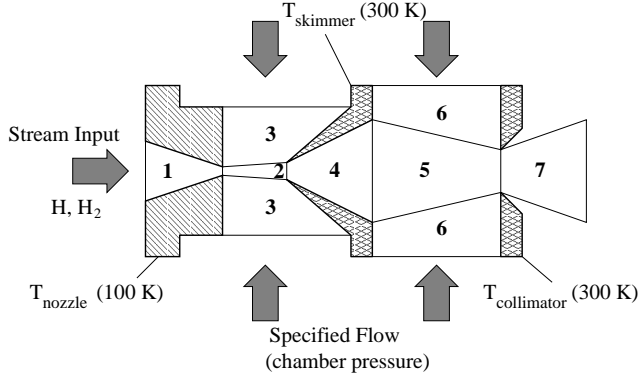}}
\caption{The geometry employed in the simulation of an expansion in the ABS.
The nozzle, skimmer, and collimator were implemented with the respective
temperatures, the space in-between is divided into regions. Due to symmetry,
only the upper half is calculated in the simulations. Specified Flows 
compensate for losses of particles at the outer boundaries of regions 3 and 6. 
There is no Specified Flow in region 7 to allow extraction of the beam 
parameters.}
\label{geopic}  
\end{figure}
Fig.~\ref{geopic} shows these 
elements together with the regions into which the flow field is divided. These 
regions are divided into the small cells mentioned above. Additionally, the 
applied input flows are indicated. The main flow is the hydrogen flow through 
the cooled nozzle. In addition flow losses have to be compensated.
So-called Specified Flows, have to 
be included to simulate the chamber pressure since particles disappear which 
pass the outer boundaries of the regions 3 and 6. The parameters of the 
gas, to be calculated for every cell of these regions, are collected in the 
Appendix.

\subsection{Experimental setup and data analysis}
\begin{figure*}[t]
\centerline{\includegraphics[height=5.5cm]{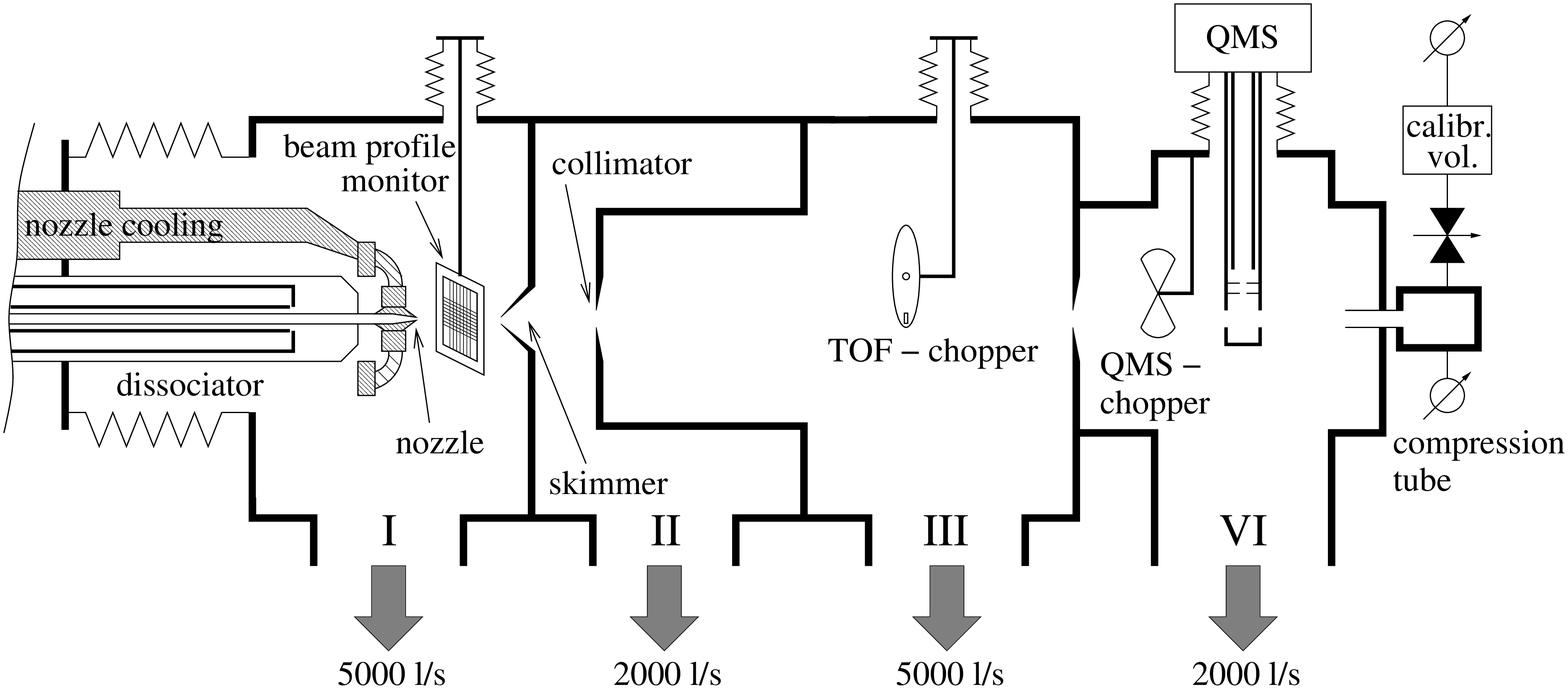}}
\caption{The atomic beam test stand with the pumping stages I-IV. The pumping 
speeds are the nominal speeds for hydrogen. The signals from the 
quadrupole mass spectrometer are read out either with a secondary-electron 
multiplier or a Faraday cup.}
\label{testst}  
\end{figure*}

An atomic beam test stand (ABT) had been set up~\cite{Koch+Steffens_1999} and 
equipped with several diagnostic devices (Fig.~\ref{testst}).
It consists of a four-stage differentially pumped vacuum system with a nominal 
pumping speed of 14000\,$l$/s. A dissociator mounted on the first chamber 
produces atomic hydrogen or deuterium. Two types of dissociators were used: 
(i) a radio frequency dissociator (RFD) \cite{Stock_et_al_Koeln_1995},
consisting of a LC circuit as a field applicator and a water cooled Pyrex
discharge tube, and (ii) a microwave dissociator 
(MWD)~\cite{Koch+Steffens_1999}, based on a plasma source which couples a 
2.45\,GHz surface wave to the discharge in an air-cooled Pyrex glass tube. 
The H (D) gas expands through a nozzle, cooled by a 120\,W closed-cycle 
helium cold head. Together with a skimmer and a collimator a high-brightness  
beam is formed. This beam is analyzed by means of several devices. A 
calibrated compression tube is used to perform absolute measurements of the 
beam 
intensity. The degree of dissociation could be determined with the quadrupole 
mass spectro\-meter (QMS). A chopper, combined with a lock-in amplifier, was 
used to separate the beam signal from the residual gas background. The 
intensity-related degree of dissociation $\alpha$, defined as 
\begin{equation}
\alpha = \frac{S^*_{\rm a}}{S_{\rm a}^* + 2\;\kappa_{\rm ion}\;
\kappa_{\rm det}\;\kappa_{\rm v}\;S_{\rm m}},
\label{aneueq}
\end{equation}
was determined from the QMS signals $S_{\rm a}$ and $S_{\rm m}$ for atoms and 
molecules, respectively. In this equation 
$S^*_{\rm a}=S_{\rm a} - \delta^{\rm diss} S_{\rm m}$ 
is the atomic signal corrected for
dissociative ionization in the QMS with $\delta^{\rm diss}=0.0125$ (0.0085) for
hydrogen (deuterium) \cite{Koch+Steffens_1999}. Furthermore, 
$\kappa_{\rm ion}=\sigma_{\rm a}^{\rm ion}/\sigma_{\rm m}^{\rm ion}=
0.64\pm0.04$
\cite{von_Engel} is the ratio of the ionization cross sections and
$\kappa_{\rm{det}}=w_{\rm a}^{\rm det}/w_{\rm m}^{\rm det}$ is the ratio 
of the detection 
probabilities of the QMS either with the secondary-electron multiplier (SEM)  
or with the Faraday cup (FC). Since one can assume 
$w_{\rm a}^{\rm FC}=w_{\rm m}^{\rm FC}$ and therefore 
$\kappa_{\rm{FC}}=1$, $\kappa_{\rm SEM}$  can be determined from the ratio
\begin{equation}
\kappa_{\rm SEM} = \frac{(S_{\rm a}^*/S_{\rm m})_{\rm SEM}}
{(S_{\rm a}^*/S_{\rm m})_{\rm FC}}\;.
\end{equation}
by measuring alternately the atomic and molecular QMS signals with the 
Faraday cup and the secondary-electron multiplier. With an accuracy 
of about $1\,\%$, $\kappa_{\rm SEM}$ was determined as 0.78~...~0.91 depending 
on the voltage, applied to the SEM. Finally, 
$\kappa_{\rm v}=v_{\rm a}/v_{\rm m}$ regards for the different dwell times of 
the atoms and molecules in the ionization volume of the QMS.

\subsubsection{Velocity analysis}
\label{velokap}
The velocity distribution of the particles in the atomic beam is determined
with the time-of-flight (TOF) method. A fast chopper cuts 
a small bunch of particles out of the beam, and their arrival time at the QMS
is measured. Since the particles have different velocities, a TOF distribution 
$F(t)$ is measured. To keep the influence of the opening function of the 
chopper small, the rotational frequency of the motor had to be high 
and stable. For this reason a special stepping motor was used, installed in 
chamber III of the setup (Fig.~\ref{testst}). A light-barrier signal defined 
the zero point of the TOF distribution and triggered the oscilloscope used to 
store the QMS signals $S(t)$.
 
The relation between the measured TOF distribution $F(t)$ and the velocity 
distribution $f(v)$ of Eq.~\ref{maxeq} with $v=l_{\rm cq}/t$ is
\begin{equation} 
F(t) \propto \frac{1}{t^2} f\left(\frac{l_{\rm cq}}{t}\right),
\end{equation}  
where $l_{\rm cq}$ is the distance between chopper and QMS. The measured 
signal distribution, corrected for an offset $p_0^*$, results from the 
convolution of the TOF function $F(t)$, the opening function of the chopper 
wheel $A(t)$, and the response function of the electronics $K(t)$ as
\begin{equation} 
S(t)-p_0^*= \int_0^t K(t-\lambda) \int_0^\lambda F(\lambda-\psi)A(\psi) 
{\rm d}\psi{\rm d}\lambda .
\end{equation} 
The opening function $A(t)$ was measured at a low rotational frequency of the 
chopper wheel. At 300\,Hz, used in the experiments, for the slit width of 
2\,mm and the tangential slit velocity of 75\,m/s, a good approximation is the 
half period of $A(t)=A_0\sin(\pi t/T)$ with $A_0=0.278$ and 
$T=0.055\pm0.004$\,ms. In the response function of the electronics, 
$K(t)=(1/\tau_{\rm e})\exp{(-t/\tau_{\rm e})}$, 
$\tau_{\rm e}=0.21\pm0.01$\,ms is calculated from the input resistance 
of the oscilloscope  and the cable capacity.

\begin{figure}
\centerline{\includegraphics[height=4.5cm,angle=0]{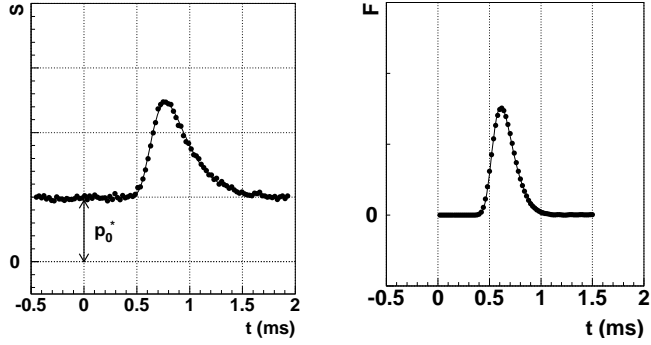}}
\caption{Example of the measured signal distributions with the offset $p_0^*$ 
and the fitted function $S(t)-p_0^*$ of Eqn.~\ref{fitfunction_1} 
(left-hand side) and the TOF distribution $F(t)$ (right-hand side), resulting 
from deconvolution, with the fitted function of Eqn.~\ref{fitfunction_2}.}
\label{velana}  
\end{figure}

The measured signal distributions were fitted by functions
\begin{equation} 
S(t)=p_0^*+\frac{p_1^*}{t^2} \exp\left\{-\frac{m}{2k_{\rm B}p_2^*}
\left(\frac{l_{\rm cq}}{t} - p_3^*\right)^2\right\} .
\label{fitfunction_1}
\end{equation}
The functions $S(t)-p_0^*$ then were de-convoluted, following the method of 
Ref.~\cite{Young_1975}, by calculating
\begin{eqnarray}
F(t_n)=\frac{\pi}{T}\sum_{i=0}^{\infty}(-1)^{i}
           \bigg[S(\lambda)+\tau_{\rm e}\frac{{\rm d}S(\lambda)}
{{\rm d}\lambda}\bigg]_{\lambda=t_n-iT}\nonumber\\
      +\frac{T}{\pi}\sum_{i=0}^{\infty}(-1)^{i}
           \bigg[\frac{{\rm d}^2S(\lambda)}{{\rm d}\lambda^2}
      +\tau_{\rm e}\frac{{\rm d}^3S(\lambda)}
{{\rm d}\lambda^3}\bigg]_{\lambda=t_n-iT}
\end{eqnarray}
for an appropriate sequence of times $t_n$, covering the time range of the TOF 
measurement. The obtained distributions $F(t_n)$ then were fitted by the TOF 
functions
\begin{equation} 
F(t) = \frac{p_1}{t^2} \exp\left\{-\frac{m}{2k_{\rm B}p_2}
\left(\frac{l_{\rm cq}}{t} - p_3\right)^2\right\}.
\label{fitfunction_2}
\end{equation} 
The fits yield $p_1$ as a scaling factor, $p_2$ according to Eqn.~\ref{maxeq} 
yields the translational beam temperature $T_{\rm tr,x}$, and $p_3$ is equal 
to the mean velocity $v_{\rm x}$ in beam direction, i.e., the beam parameters 
to be determined. Fig.~\ref{velana} as an example shows a measured signal 
distribution and the derived TOF function.

\subsubsection{Beam-profile monitor}
\label{monikap}
\begin{figure}
\centerline{{\includegraphics[height=5.5cm,angle=0]{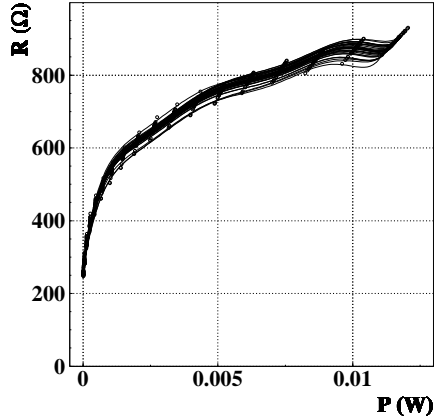}}}
\caption{The response functions of the 32 wires of the beam-profile monitor, 
measured by loading the wires with sequences of currents and measuring the 
voltages.}
\label{drahtei}  
\end{figure}

A beam-profile monitor~\cite{Vassiliev_et_al_PST99} was used to measure 
intensity profiles of the atomic hydrogen or deuterium beam. A frame, carrying 
32 gold-plated tungsten wires of diameter $d_{\rm w}=5\,\mu$m (16 wires each 
in horizontal and vertical direction) was placed between nozzle and skimmer. 
At a wire spacing of 2\,mm, the wires covered the full cross section of the 
beam. Beam atoms, which hit a wire, get sticked at the gold cover with a 
probability $w_{\rm s}^{\rm Au}=0.5 \pm 0.1$~\cite{Winkler_1998}, i.e., 
they are not immediately re-emitted. If sticked, the atoms recombine with 
atoms of the surface-covering atomic layer to form molecules, which then are 
emitted due to their much lower binding energy. The recombination probability 
$w_{\rm r}$ of sticked atoms thus can be assumed as 1.0. The recombination 
energy $E_{\rm r} = 4.476$\,eV per H$_2$ molecule~\cite{AIP_Handbook_1963} 
leads to a local heating and a differential increase d$R$ of the wire 
resistance. For a wire of length $L$ extending in $z$ direction,
${\rm d}R/{\rm d}z$ is a function $f({\rm d}P/{\rm d}z)$ of the differential 
energy deposition ${\rm d}P/{\rm d}z$. The total wire resistance is 
\begin{equation}
R=R_0 + \int\limits_{0}^{L}({\rm d}R/{\rm d}z){\rm d}z=R_0 + 
\int\limits_{0}^{L}f({\rm d}P/{\rm d}z){\rm d}z,
\end{equation}
where R$_0$ is the wire resistance without beam. It can be assumed that (i) 
the wires are homogeneous along their length and that (ii) the locally 
produced heat is spread by radiation cooling only, i.e., that the heat 
transfer along the wire can be neglected. Under these assumptions, the 
function $f$ can be determined for each wire by application of a series of 
currents $I$ and measurement of the voltages $U$. Here one has $P=U \cdot I$, 
${\rm d}P/{\rm d}z={\rm const}=P/L$, and
\begin{equation}
R=R_0 + \int\limits_{0}^{L}f(P/L){\rm d}z=R_0 + L\cdot f(P/L).
\end{equation}
The measured response functions $R_i(P)$ for the 32 wires are shown in 
Fig.~\ref{drahtei}.
With some scattering, all the curves show the same behavior, an approximately 
linear increase from $R_{i,0}$ at $P=0$ and a saturation towards higher P 
values caused by the increase of radiation cooling due to the increasing wire 
temperature.

The monitor wires are positioned at different perpendicular distances $y_i$ 
from the atomic beam axis. The changes of the total wire resistances depend on 
the total flux of atoms, hitting the wires. Thus, on the one hand the 
measurement of the set of $R_i-R_{i,0}$ yields data on the intensity 
distribution of the beam and its development with the distance $x$ from the 
nozzle exit. On the other hand, the measured data can be compared with those, 
which 
result from the DSMC calculations. These yield atomic density and velocity 
distributions, $n(x,y,z)$ and $v(x,y,z)$, which can be used to calculate the 
differential recombination-energy deposition in the wires as
\begin{equation}
{\rm d}P_i(x,y,z)=d_{\rm w} \cdot w_{\rm s}^{\rm Au}\cdot w_{\rm r}\cdot
   E_{\rm r}\cdot\frac{1}{2}n(x,y,z)\cdot v(x,y,z) {\rm d}z.
\label{dPi_wire} 
\end{equation}
With the use of Eq.~\ref{dPi_wire} and the measured response function 
$f_i({\rm d}P/{\rm d}z)=f_i(P/L)$ by summation over the wire length the 
increase of the 
wire resistance can be calculated and compared with the measured values to 
check the validity of the calculations.

\section{Simulation Results and Experimental Verification}
\subsection{Molecular hydrogen beam from a cooled nozzle}
\label{kap41}
\begin{figure}[t]
\centerline{{\includegraphics[height=5cm,angle=0]{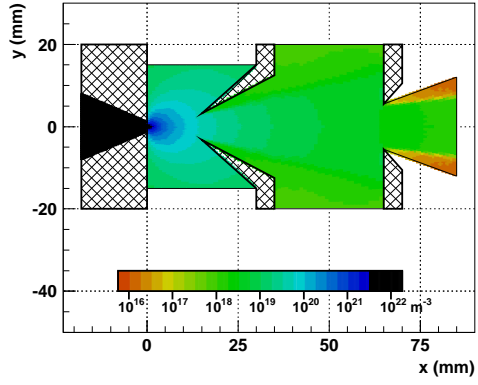}}}
\caption{The calculated particle-density distribution of a molecular hydrogen 
beam, expanding in the system of nozzle, skimmer, and collimator (diameters: 
nozzle throat 2.0\,mm, skimmer top 6.4\,mm, collimator 11\,mm; distances: 
nozzle to skimmer-top 15\,mm, skimmer-top to collimator 50\,mm; 
$T_{\rm nozzle}$ 100\,K; primary molecular gas flow 1\,mbar$l$/s).}
\label{simdich2}  
\end{figure}

First of all, a Monte Carlo simulation of an expansion of molecular hydrogen
was performed. The geometry was chosen as shown in Fig.~\ref{geopic}, the 
dimensions of the setup are given in Fig.~\ref{simdich2}. The parameters 
for the stream input were the gas temperature $T_{\rm gas}=300$\,K
and the particle density $n_0$ which was determined from the measured pressure
$p_0$ in front of the nozzle. The calculated flow rates through the nozzle, 
$Q$, were in good agreement with the measured primary gas flow. The
parameters for the Specified Flows at the outer boundaries (Fig.~\ref{geopic}) 
were obtained from the residual gas pressures $p_{\rm b}$ measured at 
the atomic beam-test stand at the respective flow rate $Q$. The calculated 
particle-density distribution of the entire simulated space is shown in 
Fig.~\ref{simdich2}.
The density near the nozzle follows a $\cos(\theta)$ distribution. Skimmer and 
collimator form a low-diverging and sharp-bound molecular beam. The density at 
the beam edge decreases by two orders of magnitude over a distance of 1\,mm at 
a beam diameter of about 15\,mm.

\begin{figure}[b]
\centerline{\includegraphics[height=8.0cm,angle=0]{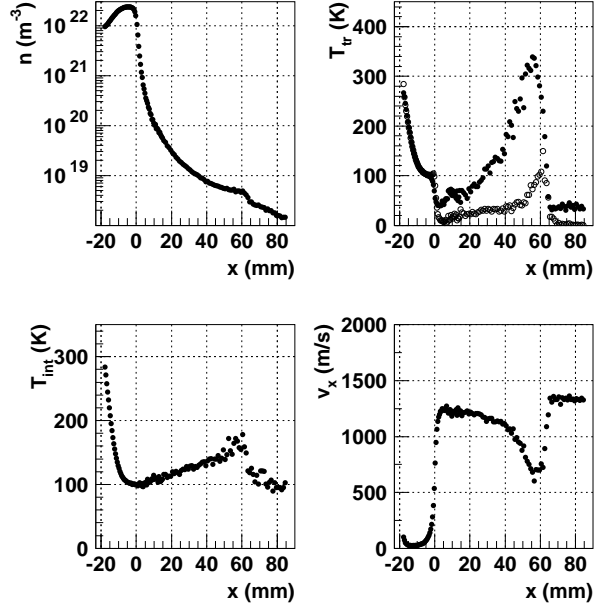}}
\caption{On-axis beam properties, calculated for a H$_2$ beam in the setup of 
Fig.~\ref{simdich2} as function of the distance to the nozzle exit 
($n$: particle density; $T_{\rm tr}$: translational temperatures describing 
the velocity distributions in $x$ (full circles), $y$ and $z$ direction 
(open circles); $T_{\rm int}$: rotational temperature; $v_{\rm x}$: mean 
particle 
velocity in $x$ direction (along the beam axis)). The pronounced variations of 
$T_{\rm tr}$, $T_{\rm int}$, and $v_{\rm x}$ between $x=2$\,mm and $x=65$\,mm 
are 
caused by the sampling of the DSMC computer code as discussed in the text.}
\label{simparh2}  
\end{figure}

Fig.~\ref{simparh2} shows the calculated on-axis beam properties.
Within the nozzle ($x=-15$\,mm to 0\,mm), $T_{\rm tr}$ and $T_{\rm int}$ drop 
to the temperature of the nozzle, i.e., the gas is thermalized, the particle 
density $n$ rises, and the mean velocity $v_{\rm x}$ is low. At the exit of 
the 
nozzle (or shortly before) the expansion into the low-pressure region starts. 
The particle density decreases with increasing distance from the nozzle.

The increase and drop of $T_{\rm tr}$ and $T_{\rm int}$ as well as the drop and
rise of $v_{\rm x}$ between $x=2$\,mm and $x=65$\,mm, the position of the 
collimator, are due to the sampling by the DSMC computer code. Every data 
point of the distributions of Fig.~\ref{simparh2} results from averaging
over the molecules in the entire cell. It includes beam molecules as well as
residual gas molecules that will not pass the collimator. Genuine beam
parameters can only be extracted from the cells with $x>65$\,mm, since 
no Specified Flow is applied to region 7 (Fig.~\ref{geopic}). Regarding this 
fact, one can state that for $x>2$\,mm $T_{\rm tr}$, $T_{\rm int}$, and 
$v_{\rm x}$ essentially stay unaffected. The reason is that the collision rate 
is too low to enable thermal equilibrium between translational and internal
energy, which requests about 100 bounces per molecule. The internal
(rotational) 
temperature ``gets frozen'' in the region of the nozzle exit. Contrary to 
$T_{\rm int}$, $T_{\rm tr}$ shows a pronounced decrease between $x=0$, the 
nozzle exit, and $x\sim2$\,mm, the position of the ``freezing surface''. A few 
bounces only are required to thermalize the translational degrees of freedom.
In the same $x$ interval, the mean velocity in beam direction, $v_{\rm x}$, 
increases with the decreasing translational beam temperature. Thermal energy 
is converted into directional motion. The temperatures $T_{\rm tr}$ in $y$ and 
$z$ directions are lower than that in $x$ direction, because particles with 
higher
transversal velocity leave the beam center, and thus the temperature on the 
axis drops. This effect may be denoted as ``geometrical cooling''.

In the upper part of Tab.~\ref{vgltab}, the values for the mean velocity 
$v_{\rm x}$ and the translation temperature $T_{\rm tr,x}$ are compared as 
they result from measurement and DSMC calculation, both performed under the 
boundary conditions given in Fig.~\ref{simdich2}.
\begin{table}
\caption{Comparison of the calculated and measured parameters of (i) a pure 
molecular and (ii) a partly dissociated hydrogen beam. The errors of the
measurements are statistical and systematical, resulting from those in the
determination of $T$ and $\tau_e$ (section \ref{velokap}), respectively. 
The asymmetric systematical errors of the simulation results originate
from discrepancies of 
the simulation distribution and the Maxwellian distribution used for
the analysis of the experimental data \cite{stancari2}. The boundary 
conditions are given in Fig.~\ref{simdich2}. The measured mean velocity in 
beam direction, $v_{\rm x}$, and the temperature $T_{\rm tr,x}$ result from 
fits according to Eqn.~\ref{fitfunction_2}. 
$M_{\rm x}=v_{\rm x}/\sqrt{\gamma k_{\rm B} T_{\rm tr,x}/m}$ 
is the axial Mach number and $v_\infty$ is the maximum beam velocity according
to Eqn.~\ref{vinfty}. The simulation calculation of 
Refs.~\cite{stancari,stancari1} were initiated by the discrepancy encountered 
with the original parameters. The degree of dissociation (Eqn.~\ref{aneueq})
of the partly dissociated beam was $\alpha=0.63$.}
\begin{small}
\label{vgltab}
\begin{tabular}{l l l l l}
\hline\noalign{\smallskip}
 & $v_{\rm x}$\,(m/s) & $T_{\rm tr,x}$\,(K) & $M_{\rm x}$ & $v_\infty$\,(m/s)\\
\noalign{\smallskip}\hline\noalign{\smallskip}
{\bf molecular beam} & & & \\
measurement & $1274\pm8\pm13$ & $19.0\pm1.1\pm0.9$ & 3.52 & \\
simulation (original)& $1334\pm12\;^{+5}_{-15}$ & $33.3\pm1.6\;^{+0}_{-3}$ 
                             & 2.79 & 1436\\
simulation (Refs.~\cite{stancari,stancari1})& $1371\pm2\;^{+5}_{-15}$ & 
                              $19.0\pm0.2\;^{+0}_{-3}$ &  3.79 & 1436\\

\noalign{\smallskip}\hline\noalign{\smallskip}
{\bf partly dissociated beam} & & \\
{\bf atoms} & & \\
~~measurement & $1750\pm47\pm24$ & $25.7\pm4.9\pm1.5$ & 2.94 & \\
~~simulation  & $1760\pm20\;^{+6}_{-19}$ & $41.0\pm3.4\;^{+0}_{-3}$ & 2.34 
& 2031\\
{\bf molecules} & & \\
~~measurement & $1579\pm51\pm17$ & $23.7\pm7.1\pm1.0$ & 3.91 & \\
~~simulation  & $1590\pm33\;^{+6}_{-17}$ & $44.0\pm4.3\;^{+0}_{-3}$ & 2.89 
& 1436\\
\noalign{\smallskip}\hline\noalign{\smallskip}
\end{tabular}
\end{small}
\end{table}
The calculated mean velocity is close to the measured one, but the 
resulting temperature is appreciably too high. The problem was 
studied~\cite{stancari,stancari1} and it was found that the temperature  
contrary to the mean velocity, strongly depends on the parameters of the 
collision processes used in the simulation code. It seems that certain 
parameters, chosen by the editors of the simulation code~\cite{DSMC}, are not 
fully correct for the low-temperature region. Furthermore, the influence of 
background gas in the nozzle-skimmer region over the beam properties was 
investigated~\cite{stancari,stancari1}. It could be shown that by the 
modification of parameters, essentially of scattering cross sections, the beam 
temperature can be adopted to the measured one, while the mean velocity stays 
almost unaffected (Tab.~\ref{vgltab}). 

\subsection{Partly dissociated beam from a nozzle}
With the use of the QMS, the TOF distributions of the atoms and molecules in 
a partly dissociated hydrogen beam were measured. The measurement was 
performed with the same setup of nozzle, skimmer, and collimator which was 
used with the pure molecular hydrogen beam. The dimensions were those given in 
Fig.~\ref{simdich2}. The degree of dissociation $\alpha$, entering the
following calculations, was between 0.50 and 0.80 depending on the
experimental conditions. TOF distributions 
were measured with nozzle temperatures $T_{\rm nozzle}$ = 70, 100, 150, and 
200\,K and primary molecular flows $Q$ between 0.5 and 6.9\,mbar$l$/s. 
The mean velocity and beam temperature for the atomic and the molecular 
fraction of the beam, measured with $T_{\rm nozzle} = 100$\,K, 
$Q=1$\,mbar$l$/s and $\alpha=0.63$ and resulting from 
the fits according to Eqn.~\ref{fitfunction_2}, together with the Mach numbers 
are given in the lower part of Tab.~\ref{vgltab}.

\begin{figure}[b]
\centerline{\includegraphics[height=4.5cm,angle=0]{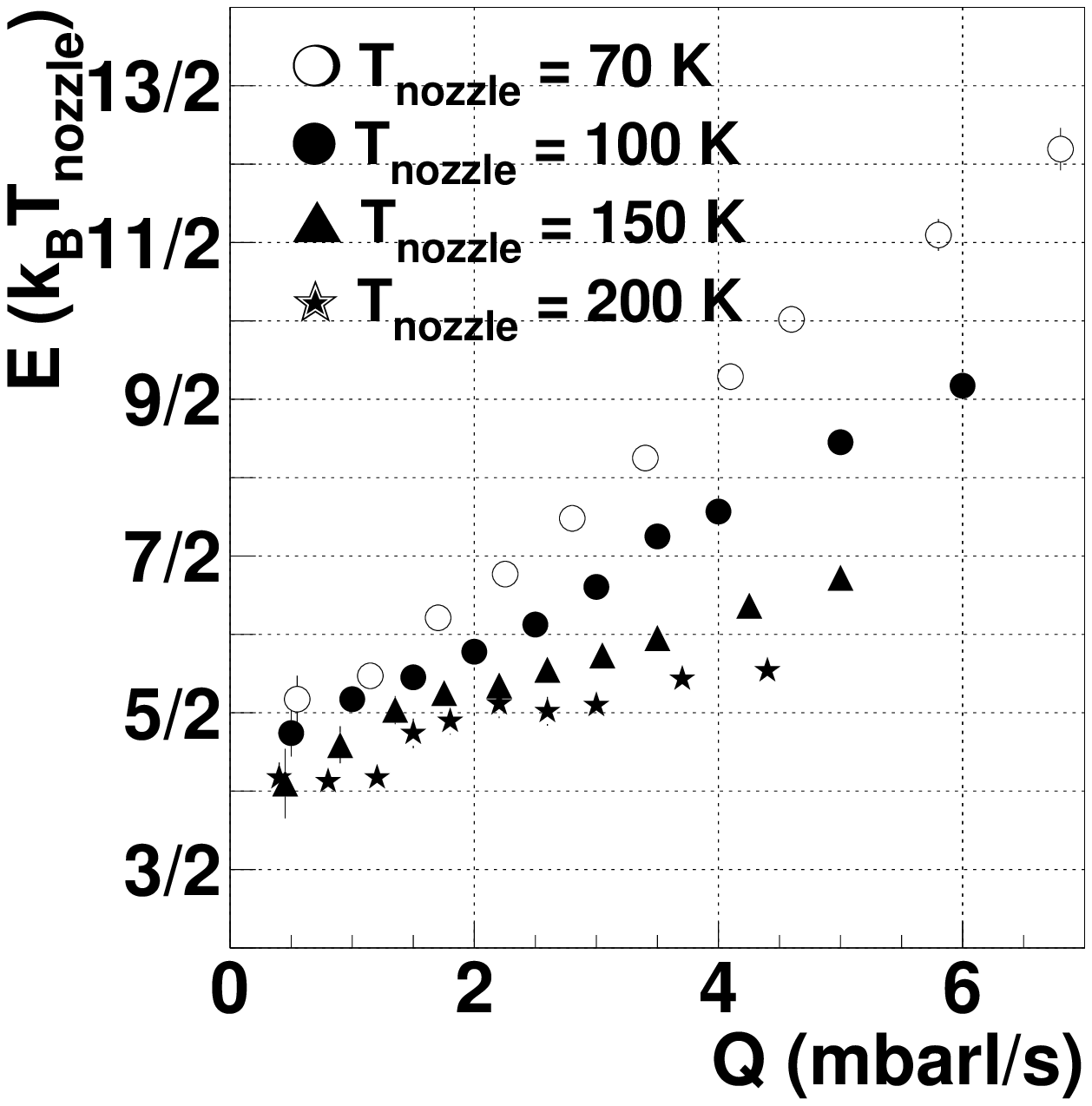}
\hspace{5mm}{\includegraphics[height=4.5cm,angle=0]{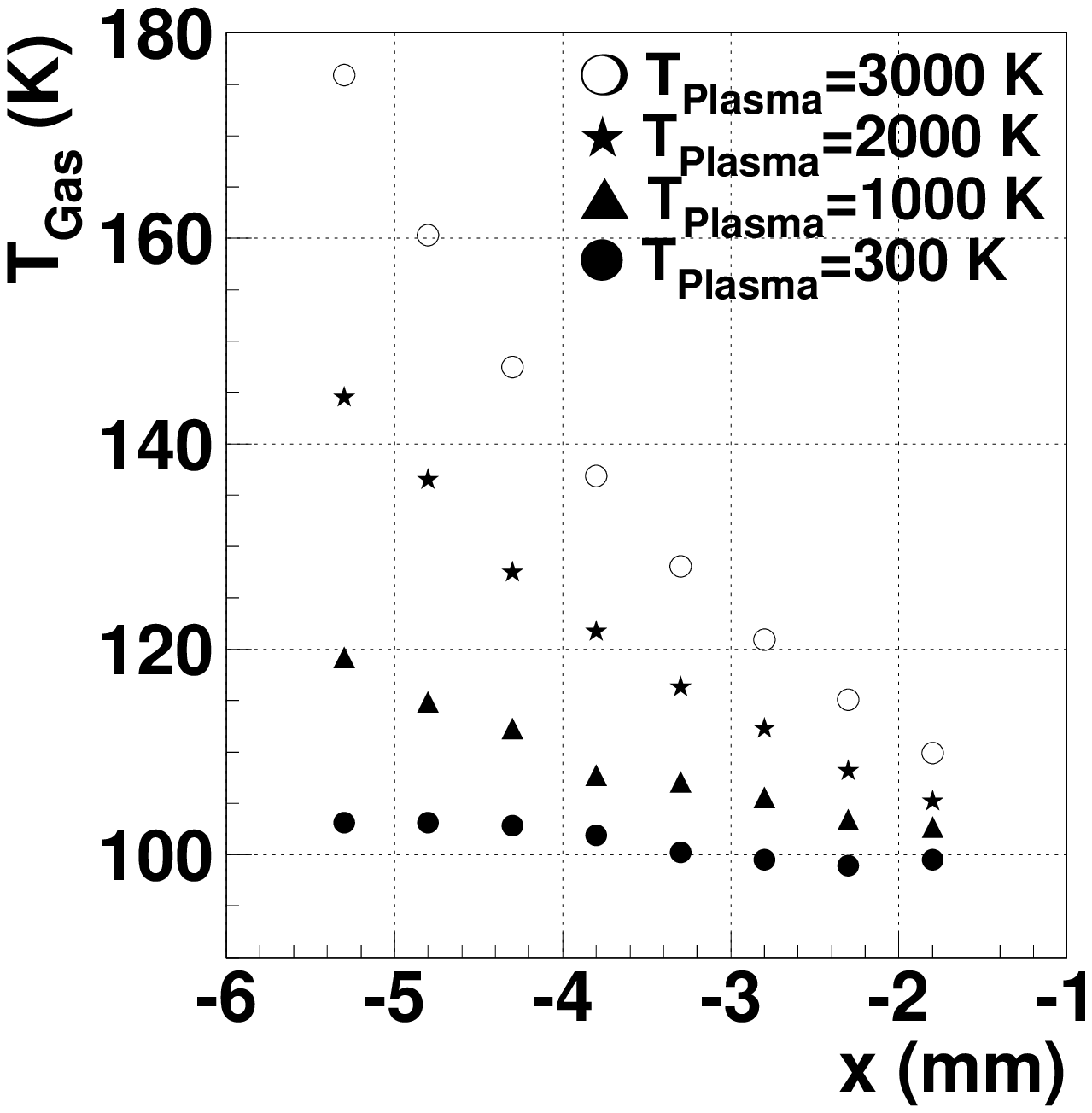}}}
\caption{Left-hand side: beam energies, as defined in Eqn.~\ref{E_definition}  
and calculated from the measured velocity distributions of H and H$_2$, as 
function of the 
primary molecular gas flow $Q$. Each of the four sets is given in units of 
$k_{\rm B} T_{\rm nozzle}$ with the respective nozzle temperature. Right-hand 
side: on-axis gas-temperature distributions in the nozzle 
($T_{\rm nozzle}=100$\,K), calculated with the DSMC code for $Q=1$\,mbar$l$/s, 
$\alpha=0.67$ and four temperatures of the plasma in the 
microwave dissociator, as function of the distance to the nozzle exit.}
\label{energy+temperature}  
\end{figure}

The temperature of the gas in the plasma of the microwave 
dissociator~\cite{Koch+Steffens_1999}, used in these measurements, may reach 
values of $T_{\rm plasma}=2000$\,K~\cite{Brown_Plasma_AIP_1993} or even 
higher. Since the plasma end is near to the nozzle entrance, one has to 
investigate, whether the gas reaches thermal equilibrium in the nozzle before 
the expansion. According to Eqn.~\ref{E_boundaries}, the beam energy (in units 
of $k_{\rm B}T_{\rm nozzle}$) is expected to lie in the range 3/2 to 5/2 for 
full temperature equilibrium of the gas in the nozzle. The left-hand part of 
Fig.~\ref{energy+temperature} shows four sets of beam energies, 
calculated from $v_{\rm x}$ and $T_{\rm tr,x}$ of both species H and H$_2$, 
as function of the primary molecular flow $Q$.

For $Q\le1$\,mbar$l$/s, the beam energies lie in the expected range 3/2 
to 5/2. For higher input flows, however, it is higher than expected. The 
discrepancy increases with decreasing nozzle temperature $T_{\rm nozzle}$ and 
increasing gas flow $Q$. This finding would be explained by the assumption 
that the temperature of the gas in the nozzle, $T_0$, is higher than 
$T_{\rm nozzle}$, i.e., that the gas does not reach the thermal equilibrium. 
For $Q=1$\,mbar$l$/s and $T_{\rm nozzle}$, this explanation was studied by 
DSMC calculations. The right-hand part of Fig.~\ref{energy+temperature} shows 
the calculated on-axis temperature of the gas in the nozzle cone
($T_{\rm nozzle}=100$\,K) as a function of distance 
to the nozzle exit for four temperatures $T_{\rm Plasma}$ of the gas entering 
the nozzle. For $T_{\rm Plasma}=300$\,K, the thermalization is complete, while 
for  higher $T_{\rm Plasma}$ the thermalization is incomplete, 
$T_0>T_{\rm nozzle}$, and hence the beam energy exceeds 
$5/2\,k_{\rm B}T_{\rm nozzle}$. Because the exact gas temperature in the 
plasma is not known, in the further DSMC calculations $T_{\rm plasma}=3000$\,K 
was used as the temperature of the gas, entering the nozzle.

\begin{figure}
\centerline{\includegraphics[height=5.cm,angle=0]{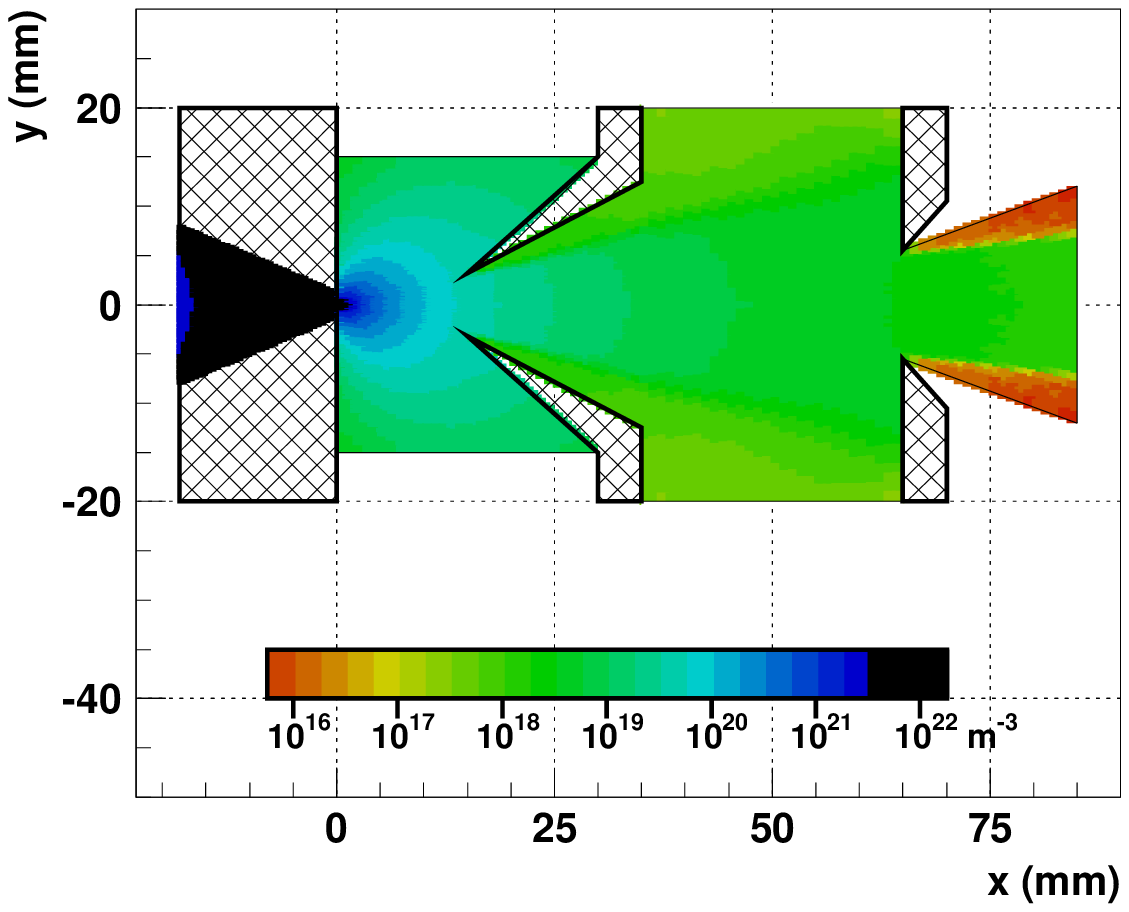}}
\vspace*{0.5cm}
\centerline{\includegraphics[height=5.cm,angle=0]{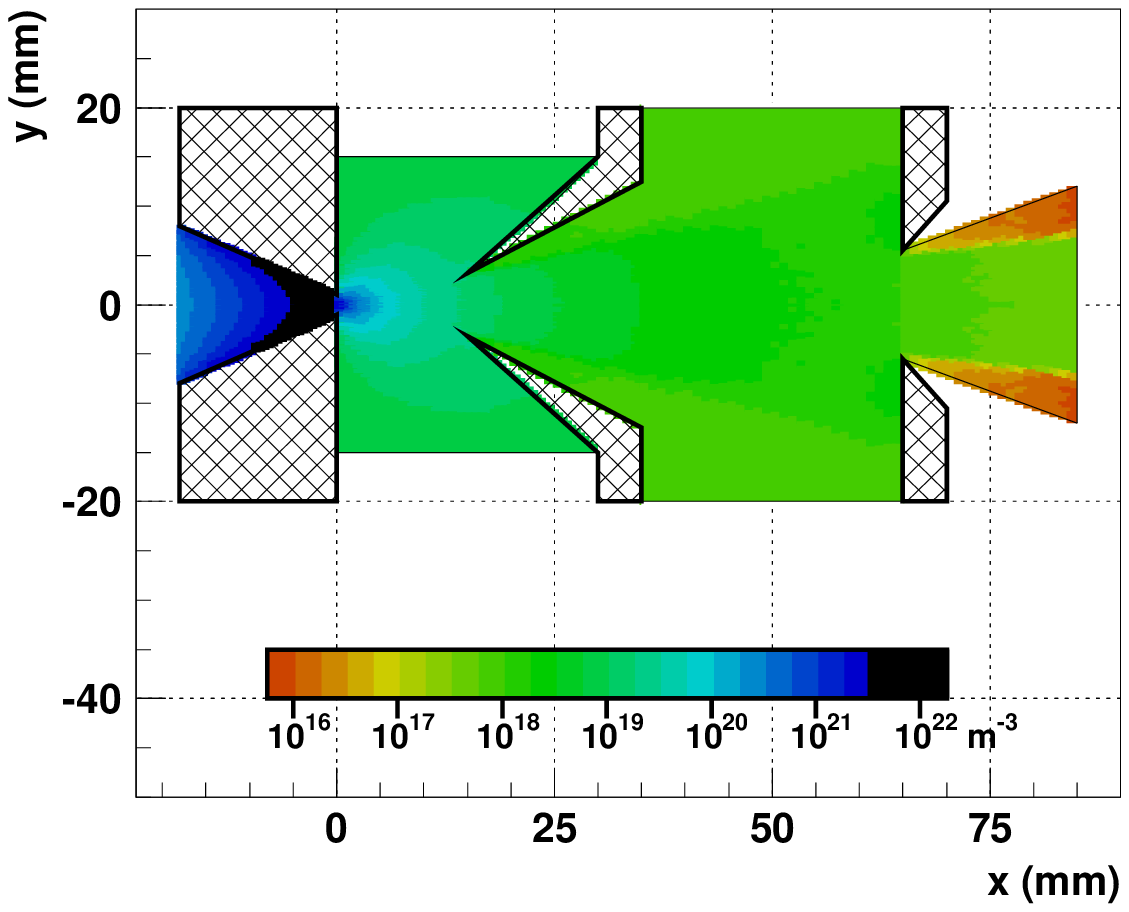}}
\caption{The calculated density distribution of the atomic (upper plot) and 
molecular (lower plot) fraction of a partly dissociated hydrogen beam 
($\alpha=0.63$) for $T_{\rm nozzle}=100$\,K and $Q=1$\,mbar$l$/s.}
\label{simdich1}  
\end{figure}

The calculated density distribution of the atomic and molecular fraction are 
shown in Fig.~\ref{simdich1}.
As for the pure molecular beam, low-diverging atomic and molecular beams with 
sharp boundaries are obtained, formed by skimmer and collimator. The 
distributions of the on-axis beam properties (particle density, translational 
and internal temperature, and mean velocity) are similar to those shown in 
Fig.~\ref{simparh2} for the pure molecular beam. The calculated mean 
velocities and temperatures for the atomic and molecular fraction in 
Tab.~\ref{vgltab} are juxtaposed with the measured values. Both the measured 
and calculated values of the mean molecular velocity are in good agreement.
They show 
that in the expansion of a gas mixture the heavier species (molecules) are 
accelerated above the theoretical limit of a pure expansion, which would yield 
$v_{\rm x,H_2}=v_{\rm x,H}/\sqrt{2}$. As for the pure molecular beam, one 
finds a deviation of the simulated from the measured temperatures, while the 
mean velocities agree. The reason for this difference is the same as discussed 
for the DSMC calculations for the molecular beam.

\subsection{Beam profiles}
\begin{figure}[t]
\centerline{\includegraphics[height=4cm,angle=0]{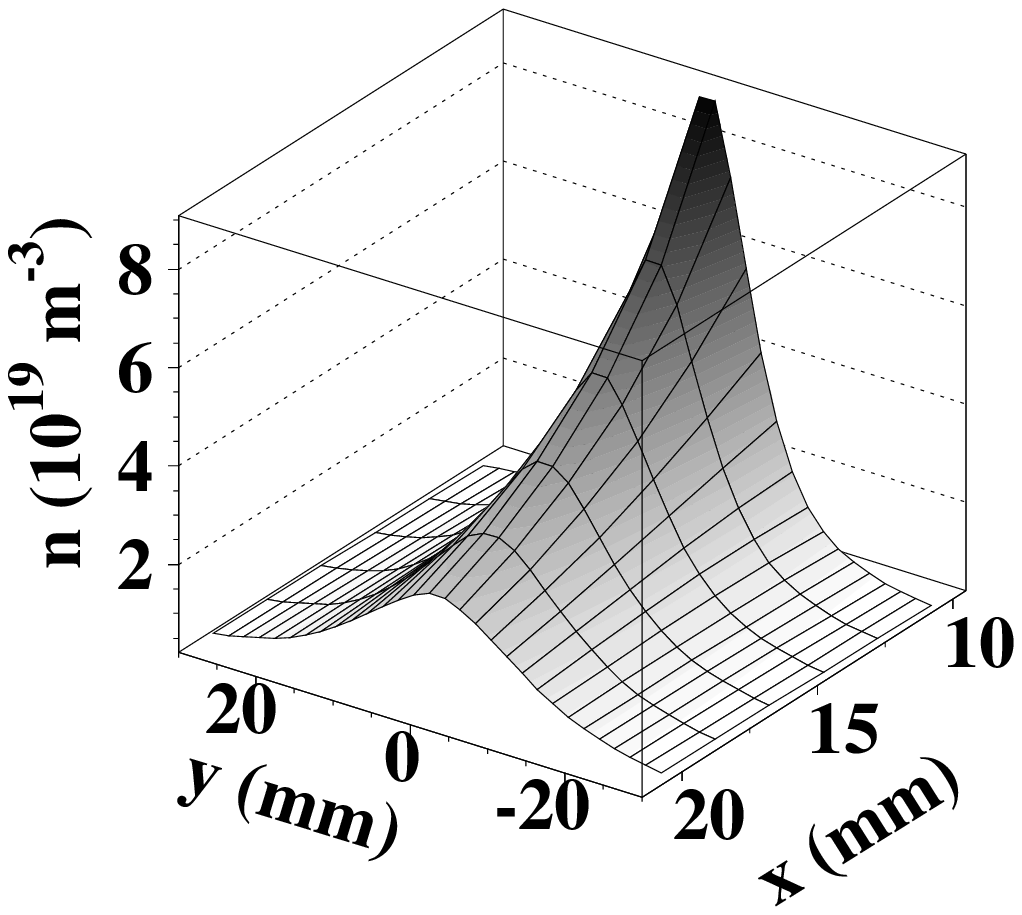}
\hspace{0.5cm}\includegraphics[height=4cm,angle=0]{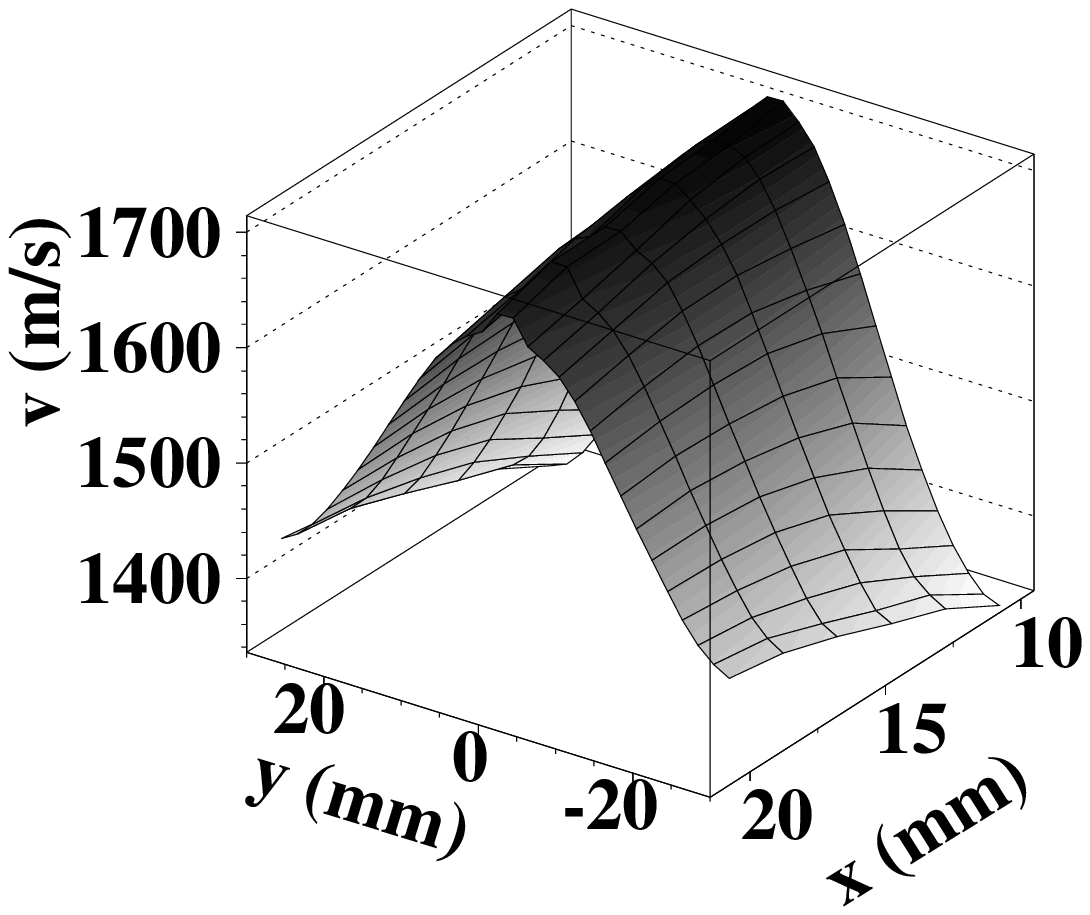}}
\caption{Calculated distributions of the atomic hydrogen density $n(x,y,z=0)$ 
(left-hand histogram) and of the mean velocity of the atoms $v(x,y,z=0)$ 
(right-hand histogram). The data of the atomic beam stem from a DSMC 
calculation of the expansion of a partly dissociated hydrogen beam 
(rf dissociator with plasma temperature $T_{\rm plasma}=3000$\,K, degree of 
dissociation $\alpha=0.76$), and nozzle temperature $T_{\rm nozzle}=100$\,K.
Because of cylindrical symmetry the dependence on $z$ at $y=0$ is identical to
the $y$ dependence at $z=0$ shown.}
\label{profdv}  
\end{figure}

The data, obtained by the DSMC calculations for a partly dissociated hydrogen 
beam, could be further compared with the 
results of beam-profile measurements, made with the beam-profile monitor  
described in the section~\ref{monikap}. The distance between nozzle and 
skimmer had to be increased from 15\,mm (Fig.~\ref{simdich2}) to 50\,mm to 
allow positioning and moving of the monitor in-between along the beam axis. 
Therefore, only the expansion from the nozzle exit into the low-pressure 
region between nozzle and skimmer was considered. The input geometry of the 
DSMC code was reduced to this region and the skimmer was assumed to be the 
transition to vacuum. The calculated density distribution $n(x,y,z)$
and velocity distribution 
$v(x,y,z)=\sqrt{v_{\rm x}^2(x,y,z)+v_{\rm y}^2(x,y,z)+v_{\rm z}^2(x,y,z)}$ 
are shown in Fig.~\ref{profdv}.
The left-hand 
histogram indicates, how the total density of particles 
decreases from the beam axis ($y=0$) to zero. Contrary to that, the mean 
velocity of the atoms, shown in the right-hand histogram, decreases only 
slightly from the on-axis value of about 1700\,m/s, comparable to those of 
Tab.~\ref{vgltab}, to about 1400\,m/s. In the covered  range of $x$, the 
density shows a distinct decrease, whereas the velocity stays more or less 
unchanged due to the fact that already the minimum $x$ position of 
10\,mm is far behind that of the freezing surface at about 2\,mm.

\begin{figure}
\centerline{\includegraphics[height=4cm,angle=0]{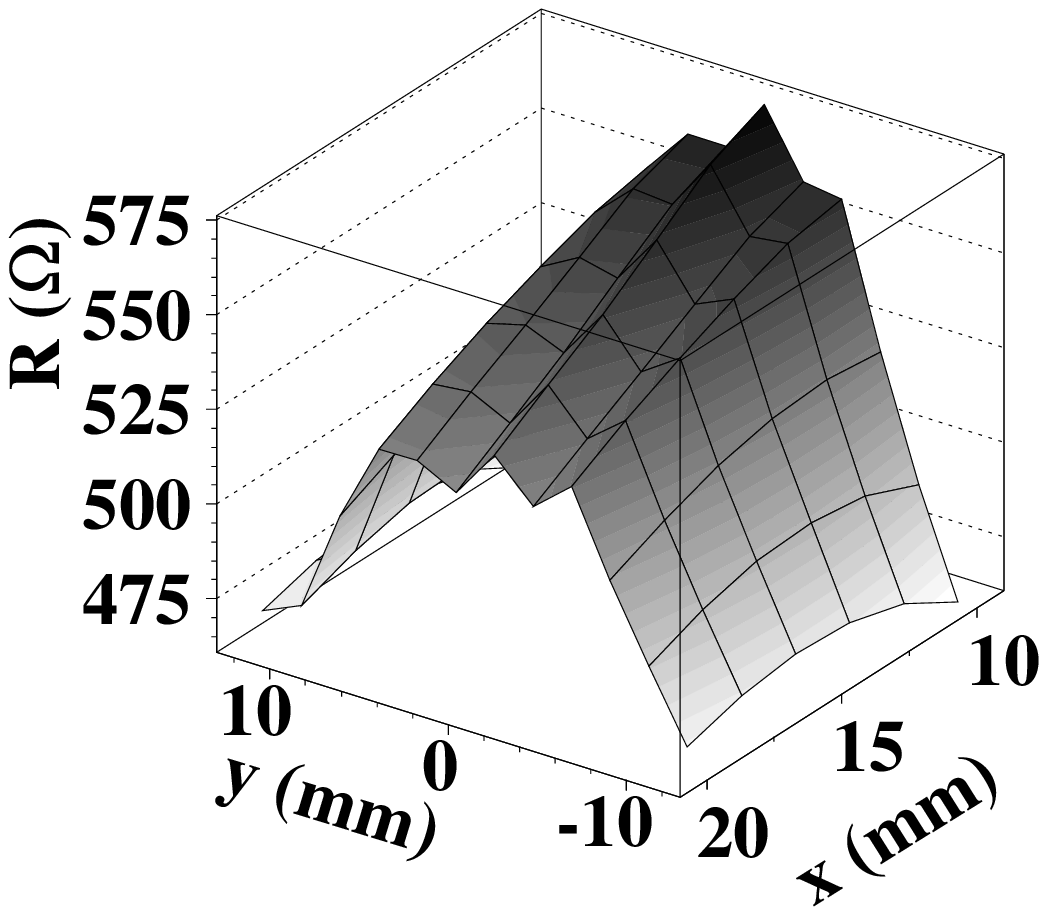}
\hspace{0.5cm}\includegraphics[height=4cm,angle=0]{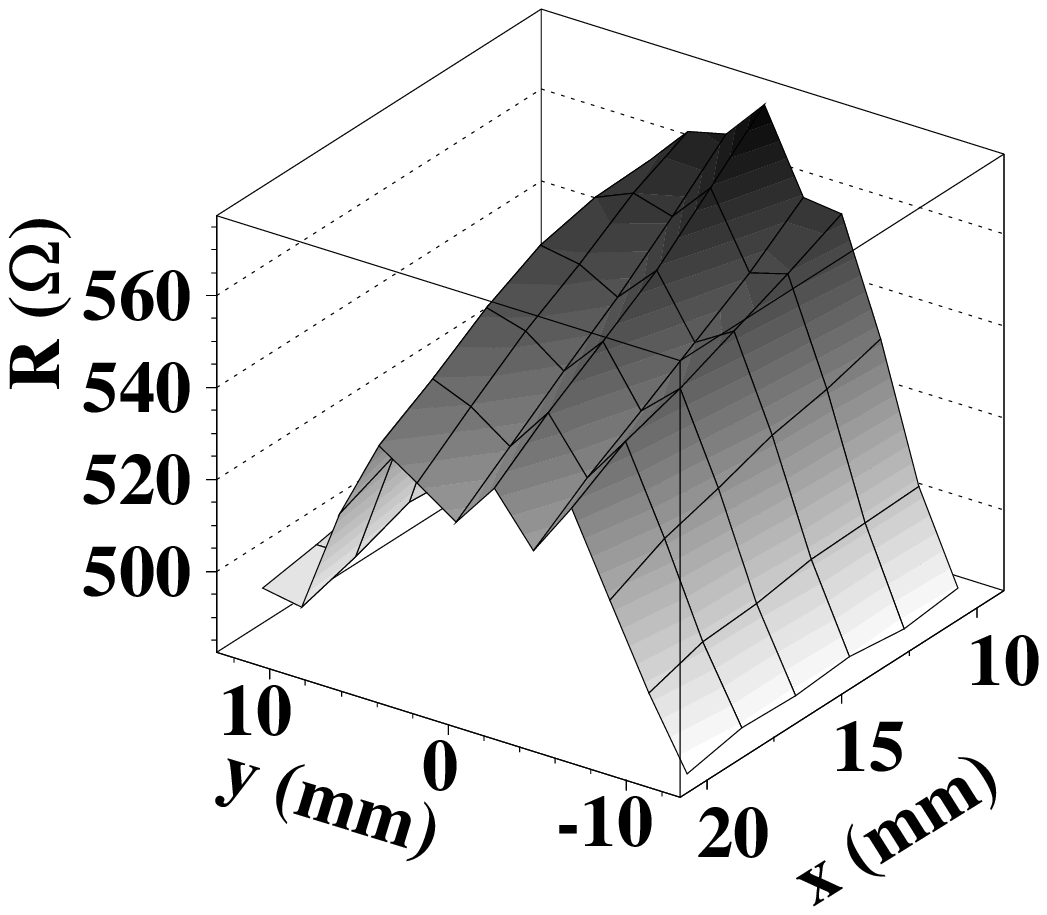}}
\caption{Calculated (left-hand part) and measured (right-hand part) 
distribution of the wire resistances (the geometry and the experimental 
parameters are those given in Figs.~\ref{simdich1} and \ref{profdv}).}
\label{profwi}  
\end{figure}

\begin{figure}
\centerline{\includegraphics[height=4cm,angle=0]{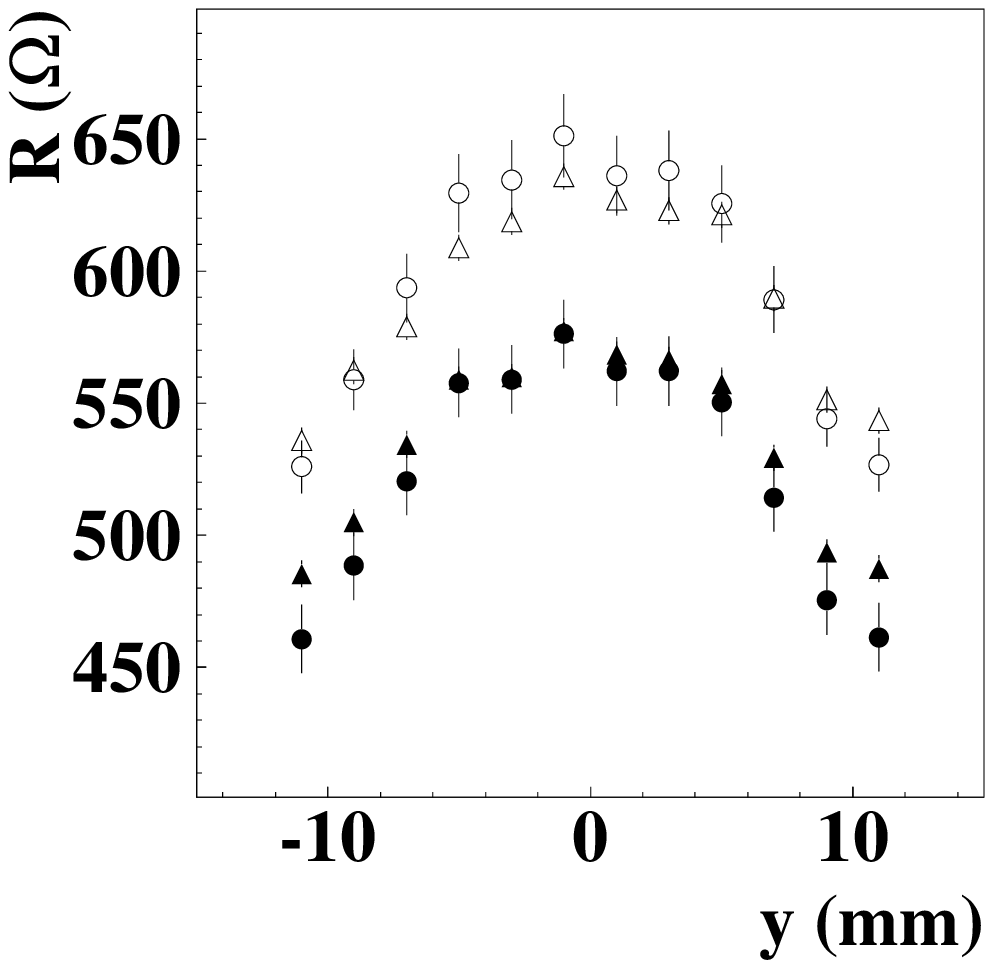}
\hspace{5mm}\includegraphics[height=4cm,angle=0]{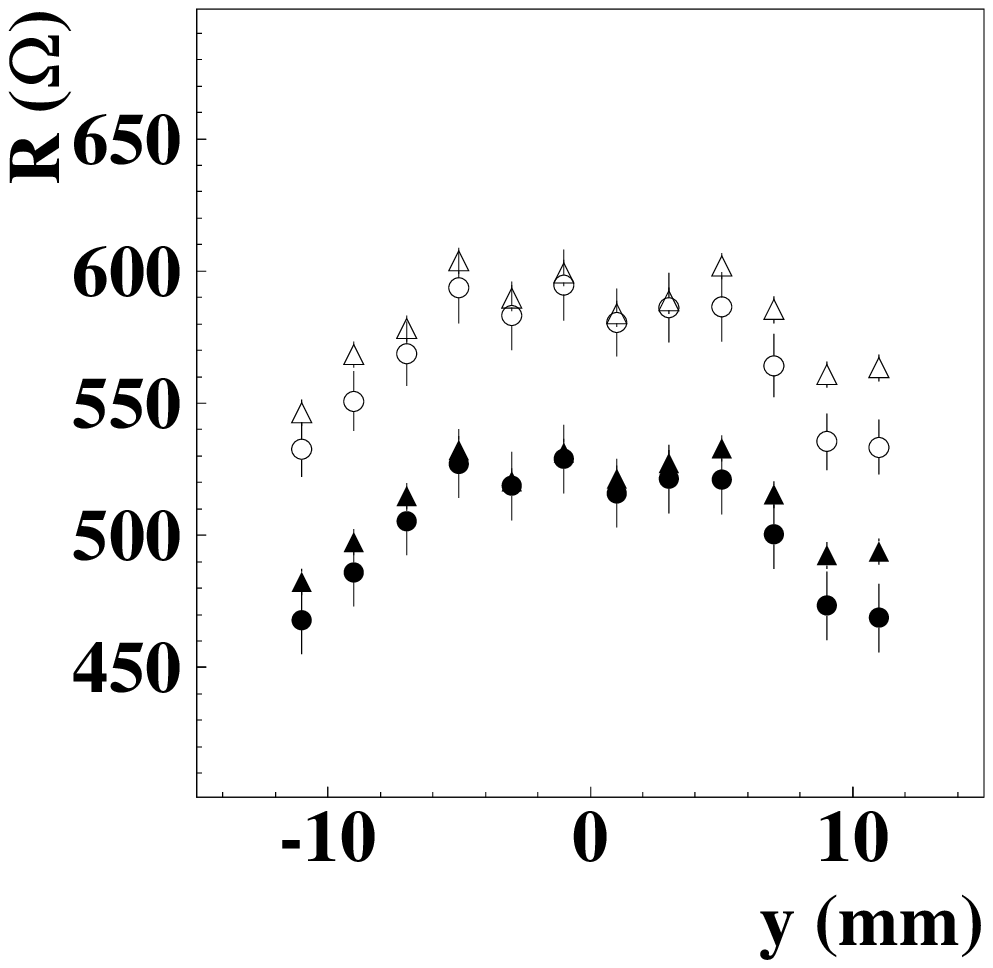}}
\caption{Comparison of the calculated (circles)and measured (triangles) 
distributions of the wire resistances at distances between wire 
plane and nozzle exit of 10\,mm (left-hand side) and 20\,mm (right-hand side). 
The lower-lying distributions (full symbols) were obtained for the geometry 
and the experimental parameters given in Figs.~\ref{simdich1} and 
\ref{profdv}, whereas the higher-lying distributions result with a primary 
molecular flow $Q$ enlarged from 1\,mbar$l$/s to 2\,mbar$l$/s.}
\label{profwg}  
\end{figure}

The beam monitor allows to measure the change of the resistance of 
each of the wires by surface recombination of the hydrogen atoms in the beam 
(cf. section~\ref{monikap}). In the used coordinate system, given in 
Fig.~\ref{expan}, each of the wires extends along the $z$ direction with a 
perpendicular distance 
$y$ to the beam axis in a plane, positioned perpendicular to the beam axis 
at distance $x$ from the nozzle exit. With the calculated $n(x,y,z)$ and 
$v(x,y,z)$, the differential heat deposition in a wire can be calculated with 
the use of Eqn.~\ref{dPi_wire}. Taking into account the measured response 
functions of Fig.~\ref{drahtei} and summation over the wire length yields 
the distribution $R(x,y)$. The density of the hydrogen atoms at the wire ends, 
$n(y=\pm25\,{\rm mm})$, is very small compared to that of the beam axis. Thus 
recombination
effects on the frame could be neglected. In Fig.~\ref{profwi}, the 
calculated distribution of the resistances is compared with the measured 
one. Both distributions are not smooth due to the variations in the
response functions of the wires. Good agreement, however, is found for the 
shape of the distributions and their absolute values. In more detail, this is 
demonstrated in Fig.~\ref{profwg}. There distributions are shown for two 
distances between wire plane and nozzle exit, $x=10$\,mm (left-hand part) and 
$x=20$\,mm (right-hand part). The lower-lying distributions result for the 
geometry of Fig.~\ref{simdich1} and the experimental parameters 
of Fig.~\ref{profdv}, i.e., they are cuts through the distributions of 
Fig.~\ref{profwi} at $x=10$\,mm and 20\,mm, 
whereas the higher-lying distributions are found for an increased primary 
molecular flow $Q=2$\,mbar$l$/s. The good agreement confirms the validity 
of the DSMC calculations.

\subsection{Hollow carrier jet}
\begin{figure}[b]
\centerline{\includegraphics[height=3.5cm,angle=0]{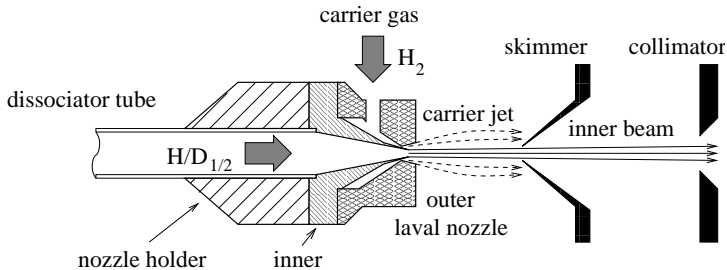}}
\caption{The nozzle and the principle of the hollow carrier jet.}
\label{carr}  
\end{figure}

The carrier jet method was proposed~\cite{Varentsov_et_al_Urbana_1997} 
to increase the phase-space density of the atomic beam and thus to reach a 
higher intensity through the collimator of the atomic beam source. An 
over-expanded carrier jet, surrounding the inner atomic beam, was predicted to 
cool and to confine the inner beam (Fig.~\ref{carr}). The mixing of the two 
gases has to be small and the carrier gas has to be removed by the skimmer and 
pumped away. The present measurements and calculations were extended to 
investigate for the first time the idea of the carrier jet. A variety of 
inner/outer gas combinations were studied, namely H/H$_2$, D/D$_2$, 
D$_2$/H$_2$, D/He, and Ar/N$_2$. The test stand 
(Fig.~\ref{testst}) had to be modified by replacing the turbo-molecular pumps 
of the first chamber by a roots-pump system with a nominal pumping speed of 
1000\,$l$/s to master the requested~\cite{Varentsov_et_al_Urbana_1997} 
high carrier-gas flows (up to 40\,mbar$l$/s). The aluminum  carrier-jet 
nozzle combines a conical inner nozzle with an outer, ring-like Laval-type 
nozzle to create an outer hollow beam surrounding an inner nozzle beam 
(Fig.~\ref{carr}).

\subsubsection{Low-mass gases}
\begin{figure}[t]
\centerline{\includegraphics[height=5cm,angle=0]{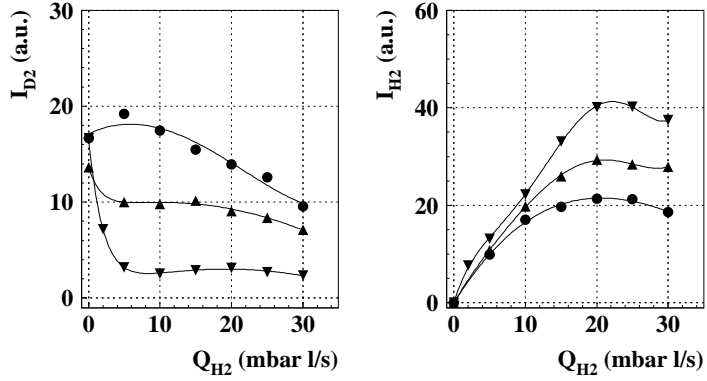}}
\caption{Measured on-axis intensities of the inner gas D$_2$ (left-hand side) 
and the carrier gas H$_2$ (right-hand side) as a function of the primary 
carrier gas flow $Q_{\rm H_2}$ for primary inner gas flows 
$Q_{\rm D_2}=1$\,mbar$l$/s (triangles down), 4\,mbar$l$/s (triangles up), and 
7\,mbar$l$/s (dots). The intensities $I_{\rm D_2}$ and $I_{\rm H_2}$ are given 
in arbitrary, but identical, units and, thus, the given values can be compared 
directly.}
\label{d2h2int}  
\end{figure}

\begin{figure}[b]
\centerline{\includegraphics[height=5cm,angle=0]{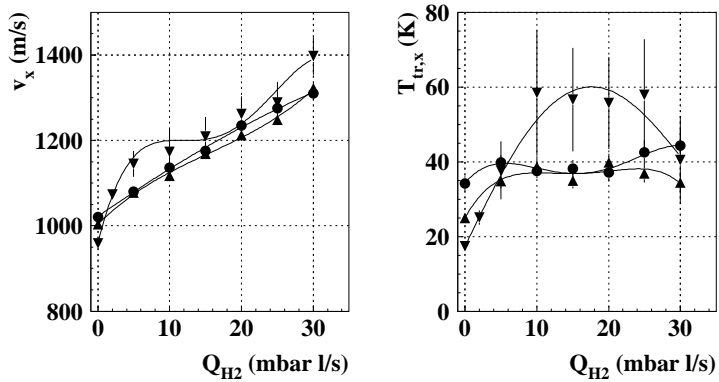}}
\caption{Mean velocity $v_{\rm x}$ and beam temperature $T_{\rm tr,x}$ of the 
inner gas D$_2$  
as a function of the primary carrier gas flow $Q_{\rm H_2}$ for primary inner 
gas flows $Q_{\rm D_2}=1$\,mbar$l$/s (triangles down), 
4\,mbar$l$/s (triangles up), and 7\,mbar$l$/s (circles).}
\label{d2h2ges}  
\end{figure}

Starting with molecular deuterium D$_2$ as inner gas and molecular hydrogen
H$_2$ as carrier gas, measurements were performed for a wide range of input 
parameters. The nozzle-throat diameter was 2\,mm and the distance between the  
nozzle exit and the skimmer was 30\,mm. The nozzle temperature was kept at 
$T_{\rm nozzle}=100$\,K. For three values 
of the inner D$_2$ flow of 1, 4, and 7\,mbar$l$/s, the outer H$_2$ flow was 
varied from 0 up to 30\,mbar$l$/s to approach the suggested 
ratio~\cite{Varentsov_et_al_Urbana_1997} of about 40. The QMS signals, which 
give the particle density in the ionization volume, were multiplied by the 
mean velocity, determined by TOF measurements, to yield the on-axis 
intensities. The results are presented in Fig.~\ref{d2h2int}. As the 
left-hand part of the figure shows, no increase of the on-axis D$_2$ beam 
intensity is found. On the contrary, especially the intensity shows a strong 
decrease at 1\,mbar$l$/s primary D$_2$ flow. The results of 
Fig.~\ref{d2h2int} were confirmed in further measurements with different 
dimensions of the outer nozzle, with different nozzle-to-skimmer distances, 
and with H/H$_2$, D/D$_2$, and D/He as inner/outer gases. The measured 
high on-axis intensities of the carrier gases indicate a pronounced mixing 
of the outer gas into the inner beam. This is obvious, too, from the beam 
parameters, mean velocity and temperature, measured for the (inner) D$_2$ 
beam as shown in Fig.~\ref{d2h2ges}. The gas of the inner beam gets 
more and more accelerated with increasing carrier-jet flow. No 
cooling effect on the inner beam was measured.

\begin{figure}[t]
\centerline{\includegraphics[height=8.8cm,angle=0]{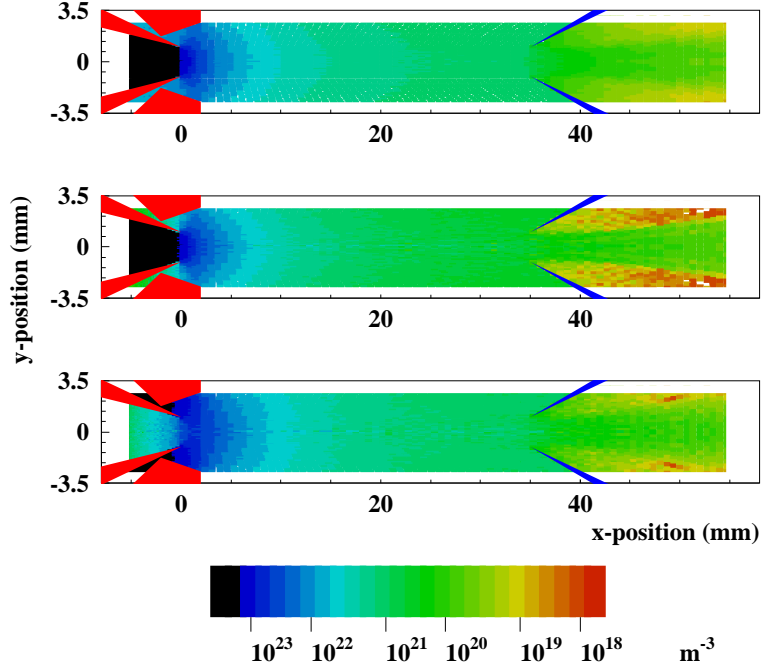}}
\caption{Calculated distributions of the density in the expansion of an inner 
D$_2$ beam without carrier jet (upper plot) and the densities of the inner, 
D$_2$, fraction (central plot) and the outer, H$_2$, fraction (lower plot) in 
the expansion with a carrier-jet. The gas flows were chosen as 
$Q_{\rm D_2}=7$\,mbar$l$/s and $Q_{\rm H_2}=21$\,mbar$l$/s.}
\label{simd2h2}  
\end{figure}

Because of the pronounced discrepancy between the measured data and the 
prediction~\cite{Varentsov_et_al_Urbana_1997}, the DSMC program was used to 
understand the mechanisms involved. The calculations were performed for a 
nozzle-throat diameter of 2\,mm, equal to that used in the measurements, 
and a 
slightly increased distance between nozzle exit and skimmer top of 35\,mm. The 
obtained density distributions are shown in Fig.~\ref{simd2h2}. There, the 
upper plot shows that of the D$_2$ beam of gas flow 
$Q_{\rm D_2}=7$\,mbar$l$/s, expanding without carrier gas. The central plot 
presents the distribution of the D$_2$ fraction ($Q_{\rm D_2}=7$\,mbar$l$/s 
from the inner nozzle as in the upper plot) in the expansion with an H$_2$ 
carrier jet ($Q_{\rm H_2}=21$\,mbar$l$/s) from the outer ring-shaped nozzle. 
The lower plot shows the density distribution of the H$_2$ carrier molecules.
No significant difference is found between the 
distributions in the upper and central plot. In agreement with the 
measurements, the comparison of the central and the lower plot shows that the 
on-axis density of the H$_2$ carrier gas even can exceed that of the D$_2$ 
beam.
A large amount of the carrier gas passes through the skimmer in contrast to
the predictions. It seems that the mixing coefficients of hydrogen/deuterium
are too high at these densities to create the carrier jet effect.
The Monte Carlo simulations again verified the experimental results. It can be
concluded that the Navier-Stokes equations, applied for the 
predictions~\cite{Varentsov_et_al_Urbana_1997} are not valid in the 
investigated flow range.

\subsubsection{Heavier gases}
\begin{figure}
\centerline{{\includegraphics[height=5cm,angle=0]{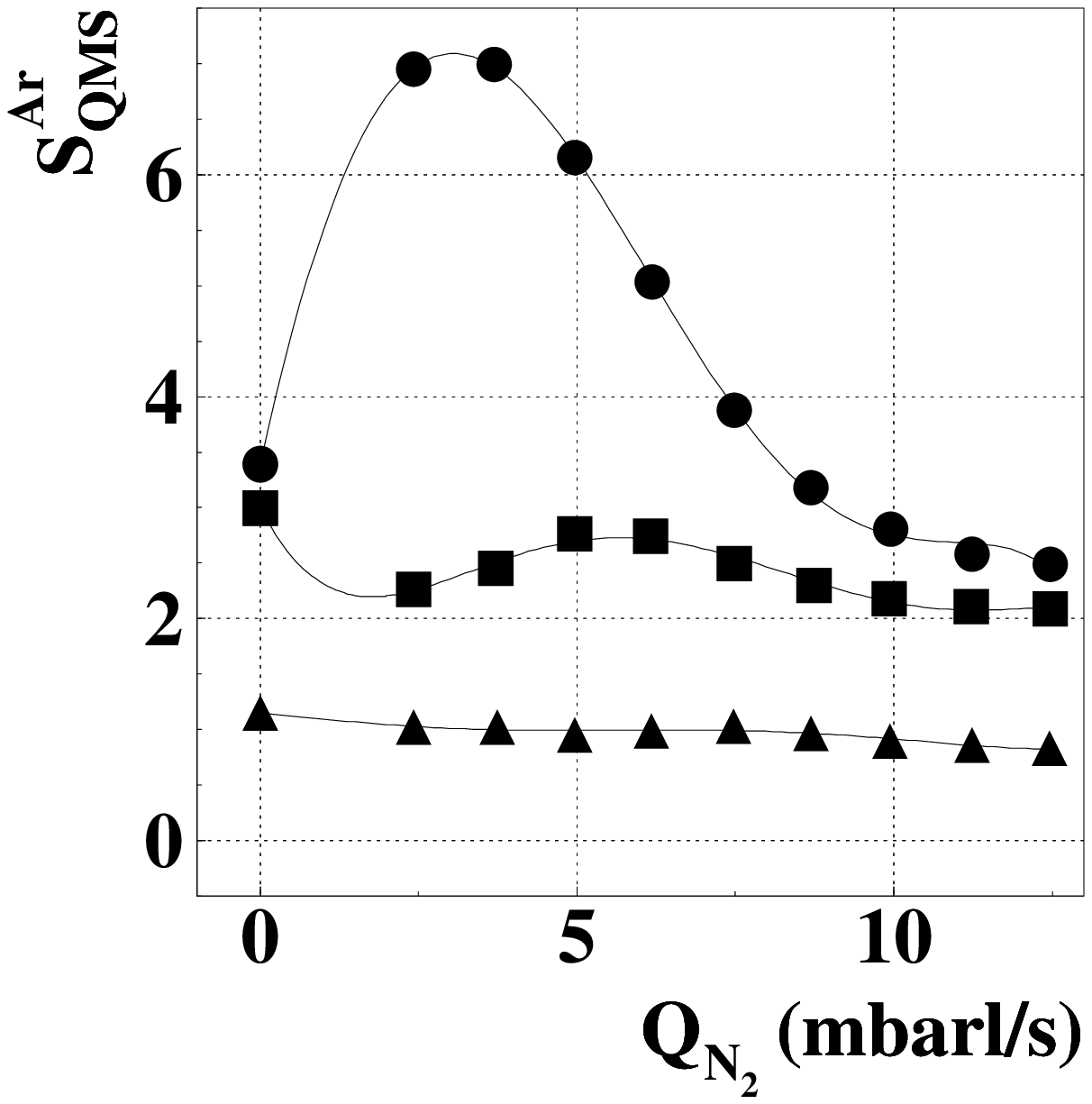}
\hspace{5mm}\includegraphics[height=5cm,angle=0]{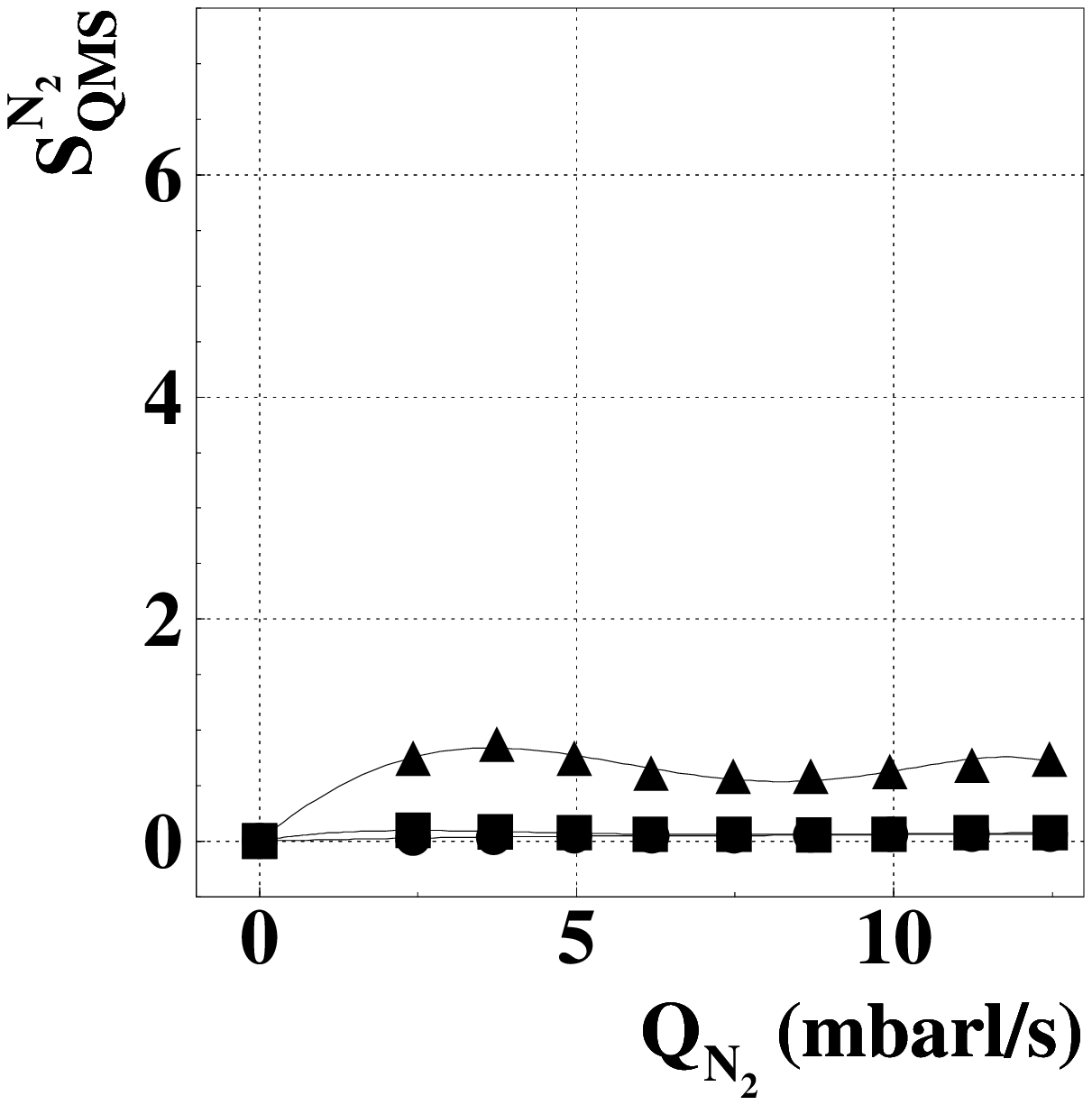}}}
\caption{The QMS signal of the inner argon gas (left) and the 
nitrogen carrier gas (right) as a function of the carrier gas flow 
$Q_{\rm N_2}$ for inner argon flows $Q_{\rm Ar}=1$\,mbar$l$/s (triangles), 
4\,mbar$l$/s (squares), and 7\,mbar$l$/s (dots). The distance between 
skimmer-top and collimator was 35\,mm and $T_{nozzle}=100$\,K.}
\label{arn2int}  
\end{figure}

\begin{figure}[b]
\centerline{\includegraphics[height=5cm,angle=0]{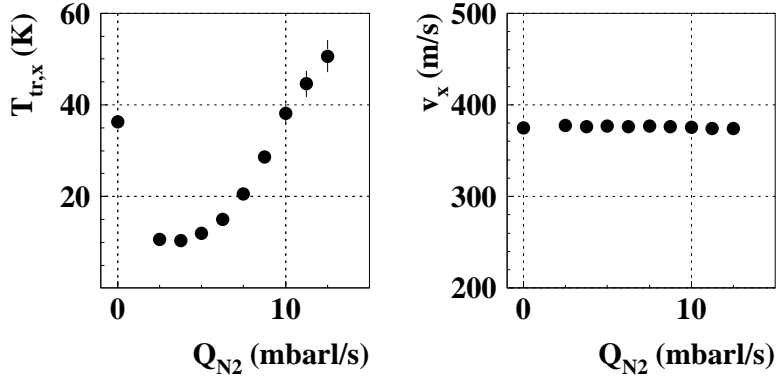}}
\caption{Beam temperature $T_{\rm tr,x}$ and mean velocity $v_{\rm x}$ of 
the inner argon gas 
as a function of the carrier gas flow rate $Q_{\rm N_2}$ for the inner 
argon flow of 7\,mbar$l$/s (distance between skimmer top and collimator 
35\,mm, $T_{\rm nozzle}=100$\,K, cf. Fig.~\ref{arn2int}).}
\label{arn2ges}  
\end{figure}

The mixing of two gases depends on their atomic or molecular diameters and 
masses. Thus an additional measurement was made using argon as inner and 
nitrogen as carrier gas. The dependences of the argon and nitrogen on-axis 
intensities on the nitrogen carrier-jet flow are found in Fig.~\ref{arn2int}.
Here, contrary to the measurements with hydrogen and 
deuterium, the QMS-signal distributions are given without multiplication by 
the velocity, because the measured mean beam velocity $v_{\rm x}$ does not 
change in the studied range of $Q_{\rm N_2}$ (see Fig.~\ref{arn2ges}). The 
distributions of Fig.~\ref{arn2int} are clearly different from those, 
measured with D$_2$ as inner and H$_2$ as carrier gas. For the highest argon 
flow of 7\,mbar$l$/s, the measured on-axis argon intensity increases by a 
factor two from the argon expansion without carrier gas. Almost no nitrogen 
was detected in the QMS at argon flows higher than 3\,mbar$l$/s.

The confirmation of the carrier-jet effect is found in the results of the
TOF measurements. Fig.~\ref{arn2ges} shows the parameters of the beam 
with inner (argon) flow $Q_{\rm Ar}=7$\,mbar$l$/s as a function of the 
nitrogen-carrier flow $Q_{\rm N_2}$. A clear cooling effect could be seen 
with a minimum around $Q_{\rm N_2}=3$\,mbar$l$/s, exactly where the
intensity maximum appears (upper curve Fig.~\ref{arn2int} left side). Thus the 
N$_2$ beam cools the Ar beam without mixing because the mean velocity stays 
constant.

\begin{figure}
\centerline{\includegraphics[height=9cm,angle=0]{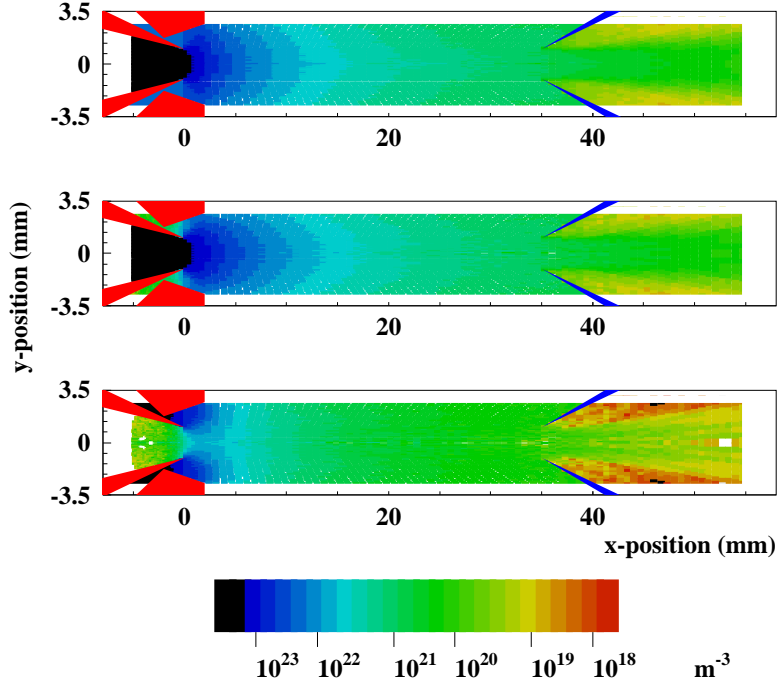}}
\caption{Calculated distributions of the density in the expansion of an inner 
argon beam without carrier jet (upper plot) and the densities of the inner, 
argon, fraction (central plot) and the outer, N$_2$, fraction (lower plot) in 
the expansion with a carrier-jet. The gas flows were chosen as 
$Q_{\rm Ar}=7$\,mbar$l$/s and $Q_{\rm N_2}=4$\,mbar$l$/s.}
\label{simarn2}  
\end{figure}

The density distribution calculated by the Monte Carlo simulation program of
the argon expansion without (upper) and with carrier jet (middle) is shown in
Fig.~\ref{simarn2}.
The on-axis argon density is increased with the carrier gas flow rate 
$Q_{\rm N_2}=4$\,mbar$l$/s compared to the carrier-free expansion. 
The flow rate 
through the skimmer increases, while the divergence of the beam is almost the 
same. The on-axis density of the nitrogen carrier gas is very small as 
well as the flow through the skimmer in contrast to that of H$_2$ 
Fig.~\ref{simd2h2}. In both cases, the results of the DSMC calculations 
are consistent with those from the measurements. The mixing of the two gases 
is restricted to the boundary layer, and it obviously is much smaller than 
for hydrogen into deuterium in that flow regime. Therefore higher densities 
in the expansion would be required to reach a carrier-jet effect for 
deuterium/hydrogen. The degree of dissociation, however, for flow rates 
$Q>7$\,mbar$l$/s is too low to achieve higher intensities by atomic beam 
sources.

\section{A further application of the DSMC method}
\begin{figure}
\centerline{\includegraphics[height=2cm,angle=0]{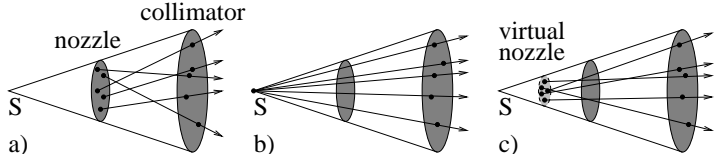}}
\caption{Models used to define the input distribution of the atoms entering 
the sextupole system in tracking calculations (a: molecular flow model, 
b: laminar flow model, c: model of a flow in the transition region).}
\label{modbild}  
\end{figure}

Apart from optimization of beam formation, the direct simulation Monte Carlo 
(DSMC) method can be applied to generate the input data for 
Monte Carlo simulations (see, e.g., Ref.~\cite{Korsch_1990}), where the
atoms are tracked through the fields of the sextupole magnets in polarized 
atomic beam sources (ABS) as it has been mentioned in the introduction. 
Position and velocity distributions of the atoms on the collimator surface are 
needed as input data to achieve a correct calculation of the output intensity 
and polarization of the ABS. In addition to the velocity distributions in beam
direction, determined by time-of-flight measurements like those of 
Ref.~\cite{Lorentz_1993}, a variety of models were used to create a sample of 
atoms passing the collimator~\cite{Braun_1995}.
\begin{itemize}
\item In the \textit{molecular flow model} the connecting line between two 
randomly distributed points, one in the nozzle exit and the other in the 
aperture of the collimator, defines the direction of the atom
(Fig.~\ref{modbild}a). 
\item The \textit{laminar flow model} uses the apex S of the cone, defined 
by the openings of nozzle and collimator, as the first point and a random 
point in the aperture of the collimator as the second one 
(Fig.~\ref{modbild}b). 
\item The generalization is the \textit{model of a flow in the transition 
region} (Fig.~\ref{modbild}c). The first point is generated on a virtual 
nozzle and the second one is a random point in the aperture of the 
collimator. A molecularity parameter is defined as 
\begin{equation}
K_{\rm mol}=\frac{\log(1-\frac{d_{\rm vn}}{d_{\rm co}})}{\log(1-\frac{d_{\rm
      pn}}{d_{\rm co}})},
\label{Kmoldef}
\end{equation}
where $d_{\rm vn}$, $d_{\rm pn}$ and $d_{\rm co}$ are the distances of the
virtual nozzle, the physical nozzle and the collimator to the point S. For 
a given nozzle-collimator distance, the choice of $K_{\rm mol}$ defines the 
position of the virtual nozzle.
\end{itemize}
For the molecular flow model $K_{\rm mol}=1$ and for the laminar flow model 
$K_{\rm mol}=0$. Furthermore, $K_{\rm mol}>1$ means that the virtual nozzle 
lies between 
the real nozzle and the collimator. To decide, which of these models yields 
a satisfactory approximation of the real distributions of the particle 
directions and velocities, the model parameters can be compared with those 
resulting from DSMC calculations.

\begin{figure}[t]
\centerline{\includegraphics[height=4cm,angle=0]{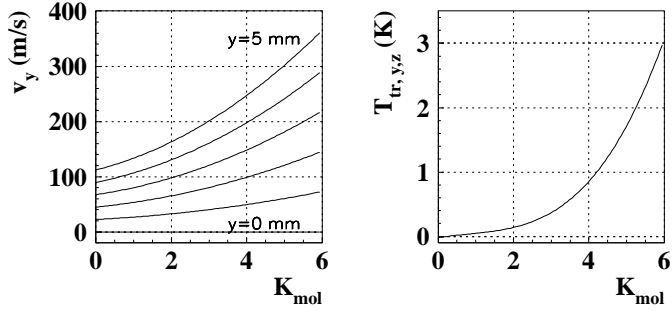}}
\caption{Mean radial velocities of the hydrogen atoms in the collimator 
plane, calculated in the framework of the models of gas expansion for 
different distances y to the beam axis (left-hand side) and non-axial beam 
temperatures (right-hand side). The curves are given as function of the 
molecularity parameter $K_{\rm mol}$.}
\label{kmolpic}  
\end{figure}

For the flow models, the unknown thermal non-axial velocity distributions can 
be approximated by a convolution of a function describing the 
nozzle-collimator 
geometry and a velocity distribution resulting from the measured mean axial 
velocity $v_{\rm x}$ and the axial beam temperature $T_{\rm tr,x}$,
\begin{small}
\begin{equation}
f(v'_j)=\int\limits_{-v'_{j,{\rm max}}}^{v'_{j,{\rm max}}}
\sqrt{1-(v/v'_{j,{\rm max}})^2} \exp\left(\frac{-m(v'_j-v)^2}{2k_{\rm B} 
T_{\rm tr,x}c_j^2}\right) 
{\rm d}v,
\label{geofolding}
\end{equation}
\end{small}
where $j=y$ and $z$ denotes the components perpendicular to the beam direction 
in the coordinate system of Fig.~\ref{expan}. The parameter $c_j$ is given by  
the nozzle-collimator geometry and the choice of the position of the virtual 
nozzle, i.e., the value of $K_{\rm mol}$. Furthermore, 
$v'_{j,{\rm max}}=c_j \cdot v_{\rm x}$ ($j=y,z$) is the maximum transverse 
thermal velocity possible at the mean axial velocity $v_{\rm x}$ and $c_j$. 
The temperatures $T_{{\rm tr},j}$ of the gas with the use of eq.~\ref{eqtemp} 
are
\begin{equation}
T_{{\rm tr},j}=\frac{m}{k_{\rm B}}\cdot\overline{v'^2_{{\rm p},j}}
   =\frac{m}{k_{\rm B}}\cdot\frac{\int\limits_{-\infty}^{\infty} 
v'^2_j f(v'_j)dv'_j}{\int\limits_{-\infty}^{\infty}f(v'_j) dv'_j}.
\label{modeltemp}
\end{equation}
The left-hand part of Fig.~\ref{kmolpic} shows the mean radial velocity 
of particles crossing the collimator aperture at different radial distances 
$y$ from the beam axis. The right-hand part shows the non-axial 
beam-temperature dependence. The curves are given as a function of the 
molecularity parameter $K_{\rm mol}$, which enters into the 
Eqns.~\ref{geofolding} and~\ref{modeltemp} via the choice of $c_j$.

\begin{figure}[t]
\centerline{\includegraphics[height=8.5cm,angle=0]{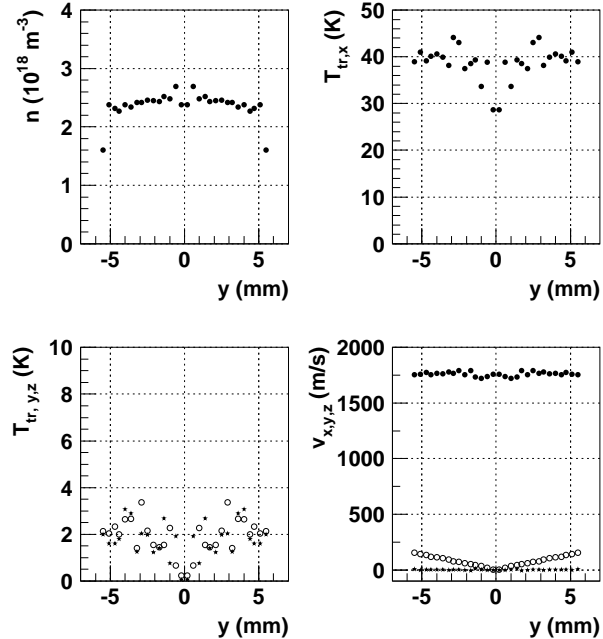}}
\caption{Beam parameters of an expanding, partly dissociated 
hydrogen beam expansion calculated with the DSMC code (primary gas flow 
$Q_{\rm H_2}=1$\,mbar$l$/s, degree of dissociation $\alpha=0.8$, 
$T_{\rm nozzle}=100$\,K, $T_{\rm plasma}=3000$\,K). In the lower plots full 
circles denote the $x$, open circles the $y$, and stars the $z$ direction.}
\label{simcoll}  
\end{figure}

The parameters, obtained in the framework of the flow models, can be 
interpreted by comparison with those from DSMC calculations, presented in 
Fig.~\ref{simcoll}. By this comparison, the validity of the three 
models
can be discussed concerning the distributions of the input data for the 
trajectory calculations. 
\begin{itemize}
\item The molecular-flow model ($K_{\rm mol}=1$) with the assumption of 
particle emission from the real nozzle allows $T_{\rm tr,z}$ to be different 
from zero, because not all the trajectories lie within planes containing the 
beam axis. The non-axial temperatures $T_{\rm tr,y}$ and $T_{\rm tr,z}$, 
resulting from the DSMC calculation 
with values around 2\,K (Fig.~\ref{simcoll}), lie higher than the small 
values of $\sim 0.1$\,K from the model approximation. The DSMC calculation, 
however, may yield too high temperature values with the standard parameters 
in the code as it was discussed in section~\ref{kap41}. Modification of the 
parameters might yield a lower temperature. The analysis of the 
density-distribution data, obtained with the DSMC code (Fig.~\ref{simdich1}) 
yields $0<K_{\rm mol}<1$, 
which supports the applicability of this model. With emission of randomly 
distributed particles from the nozzle, it is the simplest model one can think 
of. It is thought to be applicable, when high accuracy is not requested in the 
tracking calculations.
\item The laminar flow model ($K_{\rm mol}=0$) was frequently used, since the 
radial beam profile, measured behind the collimator~\cite{Korsch_1990}, was 
considered to be sufficiently consistent with that predicted by the model. 
Particle emission from a point like source on the beam axis implies that the 
non-axial beam temperatures are zero. The results of the DSMC calculations 
(Fig.~\ref{simcoll}), however, show that the transversal beam temperatures 
are different from zero (even, when they are reduced from the calculated 
$T_{\rm tr, y,z}\approx 2$\,K to 1\,K due to the reasons given above for the 
molecular flow). A temperature of 1\,K corresponds to a root mean square 
velocity of $(\overline{v'^2_{{\rm p},j}})^{1/2}=91$\,m/s. Total omission of 
those particle trajectories, not lying in planes containing the beam axis, 
means neglecting atoms with an angular velocity component and modified 
trajectories in the sextupole system due to the centrifugal force. Therefore, 
the laminar flow model is thought not to be suited to yield the appropriate 
input data. 
\item For the model of a flow in the transition region, the transversal beam 
temperatures from the DSMC calculation, reduced to 1\,K as discussed above, 
according to Fig.~\ref{kmolpic} yield $K_{\rm mol}\sim4$. For this value, 
the increase of the transversal beam velocity $v_{\rm y}$ from the beam axis 
($y=0$) to the aperture radius of the collimator ($y=5.5$\,mm) roughly match 
that of the DSMC data (Fig.~\ref{simcoll}).According to its definition, 
Eqn.~\ref{Kmoldef}, the virtual nozzle for this value is positioned between 
the real nozzle and the collimator, which might be explained with particle 
emission from the freezing zone, i.e., the zone, where the laminar flow turns 
to the molecular flow.
\end{itemize}
The molecular-flow model and the model of a flow in the transition region are 
thought to be suited to yield the input-data distributions for the trajectory 
calculations in the sextupole magnets of a polarized ABS. The latter model 
might be the superior one, when the position of the virtual nozzle can be 
fixed to the appropriate position.

\section{Summary and Conclusion}
It has been shown that the direct simulation Monte Carlo (DSMC) 
method~\cite{Bird,DSMC} is an excellent tool to describe the processes 
occurring in the expansion of light and also heavier gases in the transition 
region between laminar and molecular flow. The results of the calculations 
were confirmed by the measurements, performed at an atomic beam test stand 
(Fig.~\ref{testst},~\cite{Koch+Steffens_1999}) with the use of an 
rf dissociator~\cite{Stock_et_al_Koeln_1995}, a 
microwave~\cite{Koch+Steffens_1999} dissociator, and a novel atomic 
beam-profile monitor~\cite{Vassiliev_et_al_PST99}. The origin 
of the discrepancies between simulated and measured temperatures was 
found, the problem could be solved~\cite{stancari,stancari1} by modification 
of the appropriate input parameters of the DSMC code~\cite{DSMC}. The 
predicted carrier-jet effect~\cite{Varentsov_et_al_Urbana_1997} could not be 
observed for hydrogen and deuterium at the operational parameters of atomic 
beam sources. It was, however, observed for an argon beam surrounded by a 
molecular nitrogen-carrier jet. Both experimental findings are 
consistent with the results of DSMC calculations. The problems, occurring with 
the so far used start generators for particle-trajectory calculations through 
the sextupole magnets of polarized atomic beam sources were studied. 
Uncertainties in the 
used models remain, and the distributions from DSMC calculations are 
considered to present the most reliable start generator.

\section{Acknowledgments}
We are grateful for the support provided by the DESY management and the DESY
staff. In particular we acknowledge the help by the machine shops of MPI
Heidelberg and the University Erlangen-N\"urnberg. This work was supported by 
the German Bundes\-ministerium f\"ur Bildung, Wissenschaft, For\-schung und 
Technologie (BMBF 057ER12P(2), 06ER929I). Special thanks are due to N. Koch, 
A. Vassiliev, E.~H\"anisch, C.~Bartels, V.~Prahl, Y. Holler, K.~Rith 
and the members of the HERMES experiment. 
Especially we want to thank M.~Stancari for her studies of the effects by the
choice of the input parameters of the DSMC code.
Finally we thank H.~Seyfarth for the 
careful reading of the manuscript and fruitful suggestions. 

\newpage
\centerline{\large \bf Appendix}
The output parameters of the DSMC program~\cite{DSMC} in the two 
output files of the first column are shown in the table below. 
The $x$- and $y$-coordinate refer to the center 
of the corresponding cell. The mean velocity of species $i$ in $j$-direction 
is \mbox{$v_j^i = v_j - \Delta v^i_j$}. The overall temperature $T$ is 
calculated as~\cite{Bird}
\begin{equation}
T = (3 T_{\rm tr} + \bar\zeta_{\rm rot} T_{\rm int}^{\rm rot} +
\bar\zeta_{\rm vib} T_{\rm int}^{\rm vib})/(3+\bar\zeta_{\rm
  rot}+\bar\zeta_{\rm vib}),\nonumber
\end{equation}
where $\bar\zeta$ are the mean numbers of degrees of freedom of the species.
It has to be mentioned that in the axially symmetric case utilized here, $x$ is
the beam direction, $y$ the radial direction and $z$ the circumferential
direction, i.e., $v_{\rm z}$ is tangential to the angular velocity.
\begin{table}[h]
\begin{small}
\begin{tabular}{c l l}
\hline\noalign{\smallskip}
output file & parameter & description \\
\noalign{\smallskip}\hline\noalign{\smallskip}
ds2gf.txt & X COORD & $x$ coordinate \\
 & Y COORD & $y$ coordinate \\
(for the & DENSITY & mass density $\rho$ \\
entire & TR TEMP & kinetic temperature $T_{\rm tr}$\\
beam) & ROT TEMP & temp. of rotations $T_{\rm int}^{\rm rot}$\\
 & VIB TEMP & temp. of vibrations $T_{\rm int}^{\rm vib}$\\
 & OV TEMP & overall temperature $T$ \\
 & MACH & Mach number $M$\\
& $\left.
\begin{matrix}
\mbox{U}\qquad\qquad \\ 
\mbox{V}\qquad\qquad \\
\mbox{W}\qquad\qquad
\end{matrix}
\right\} $ &
$\begin{matrix}
\mbox{mean velocity} \\ 
v_{\rm x},\,v_{\rm y},\,v_{\rm z} \\
\mbox{of all species} 
\end{matrix} $ \\
\noalign{\smallskip}\hline
ds2gm.txt & N DENS & number density $n$ \\
 & TTX & kinetic temp. in x $T_{\rm tr,x}$\\
(for every & TTY & kinetic temp. in y $T_{\rm tr,y}$\\
species) & TTZ & kinetic temp. in z $T_{\rm tr,z}$\\
 &
$\left.
\begin{matrix}
\mbox{U DIF VEL} \\ 
\mbox{V DIF VEL} \\
\mbox{W DIF VEL}
\end{matrix}
\right\} $ &
$\begin{matrix}
\mbox{mean velocity difference} \\ 
\mbox{of a species $i$ in x, y, z} \\
\Delta v^i_{\rm x},\,\Delta v^i_{\rm y},\,\Delta v^i_{\rm z}
\end{matrix} $ \\
\noalign{\smallskip}\hline
\end{tabular}
\end{small}
\end{table}
\end{document}